\documentclass[prb,twocolumn,showpacs]{revtex4}

\usepackage{graphicx}
\usepackage{dcolumn}
\usepackage{bm}
\usepackage{color}

\def\cf#1{\ensuremath{{\rm #1}}}

\begin{document}


\title{Ab initio study of gap opening and screening effects in gated bilayer graphene}

\author{\hspace{-1.7cm} Paola Gava, Michele Lazzeri,
A. Marco Saitta, and Francesco Mauri\\[1em]
IMPMC,  CNRS, IPGP, Universit\'es Paris 6 et 7, 140 Rue de Lourmel,
75015 Paris, France
\normalsize
}

\date{\today}

\begin{abstract}

The electronic properties of doped bilayer graphene in presence of
bottom and top gates have been studied and characterized by means
of density-functional theory (DFT) calculations. Varying independently
the bottom and top gates it is possible to control separately the
total doping charge on the sample and the average external
electric field acting on the bilayer. We show that, at fixed
doping level, the band-gap at the $K$ point in the Brillouin zone
depends linearly on the average electric field, whereas the
corresponding proportionality coefficient has a nonmonotonic
dependence on doping. We find that the DFT-calculated band-gap at
$K$, for small doping levels, is roughly half of the band-gap
obtained with standard tight tinding (TB) approach. We show that this
discrepancy arises from an underestimate, in the TB model, of the
screening of the system to the external electric field. In
particular, on the basis of our DFT results we observe that, when
bilayer graphene is in presence of an external electric field,
both interlayer and intralayer screenings occur. Only the
interlayer screening is included in TB calculations, while both
screenings are fundamental for the description of the band-gap
opening. We finally provide a general scheme to obtain the full
band structure of gated bilayer graphene for an arbitrary value
of the external electric field and of doping.
\end{abstract}

\maketitle


\section{\label{sec:intro} Introduction}

Among the nanoscale forms of carbon, bilayer graphene has recently
attracted much interest.
\cite{McCann-PRL-96,Guinea-PRB-73,Ohta_science,Oostinga-Nature,McCann-PRB-74,Min-PRB-75,Castroneto_PRL,Castroneto_arXiv}
Indeed, it has been found, both theoretically and experimentally,
that in presence of an asymmetry between the two graphene layers,
generated by an external electric field, a band-gap can be opened.
This makes bilayer graphene a tunable-gap semiconductor and
therefore an exciting structure for future application in
nanoelectronics.

In particular, in the experiments of Ohta et al.
\cite{Ohta_science} bilayer graphene is synthesized on silicon
carbide (SiC) substrate, and a small $n$-type doping is acquired
by the system from the substrate. In this case, the bilayer
symmetry is broken by the dipole field created by the depletion of
charge on SiC and accumulation of charge on the bilayer.
Additional $n$ doping is induced by deposition of potassium atoms
above the bilayer. Varying the concentration of potassium atoms,
one can vary the asymmetry between the two sides of the system and
measure the electronic properties and the band-gap opening by
angle-resolved photoemission spectroscopy (ARPES).

Oostinga et al. \cite{Oostinga-Nature} used a double-gated system,
where monolayer and bilayer graphenes are placed  between two
dielectrics, which act as bottom and top gates. The double-gated
structure gives the possibility to control independently the
doping level and the perpendicular electric field acting on the
system. In this configuration, they measure the dependence of the
resistance on the temperature and on the electric field. They
observe a gate-induced insulating state in bilayer graphene which
originates from the band-gap opening between the valence and
conduction bands.

As for theoretical studies, McCann \cite{McCann-PRB-74} used a
tight binding (TB) model to study the band structure of the bilayer
graphene in presence of an energy difference between the two
layers, which determines a band-gap opening. In particular, he
considered a single gate acting on the system, and he found a
roughly linear relation of the gap with the accumulated charge $n$
on the bilayer, for $n$ values up to \cf{10 \times 10^{12}
cm^{-2}}. Min et al. \cite{Min-PRB-75} performed ab initio density-functional 
theory (DFT) calculations of undoped bilayer graphene
in a constant external electric field, using the generalized
gradient approximation (GGA) for the exchange-correlation
functional. They confirmed the general picture provided by the TB
model, although DFT screening results stronger. Moreover, Aoki et
Amawashi\cite{Aoki-solstatecomm-142} performed an ab initio DFT study
on the band structure dependence of undoped layered graphene on
the stacking and external field, using the local density
approximation (LDA) for the exchange-correlation functional. In
contrast with the GGA study of Min et al., \cite{Min-PRB-75} their
results on undoped bilayer graphene in a uniform external
electric field are in agreement with the TB ones. Again, Castro et
al. \cite{Castroneto_PRL,Castroneto_arXiv} showed experimentally,
by measuring the Hall conductivity and the cyclotron mass of
biased bilayer graphene, and theoretically, using TB methods, that
a band-gap between the valence and conduction bands can be tuned
by an applied electric field.

Other theoretical DFT studies have been devoted to the
understanding of the band structure dependence on the stacking
geometry, \cite{Latil-PRL} on the presence of adsorbed molecules,\cite{Ribeiro}
 and to the analysis of the distribution
of the induced charge densities. \cite{Yu-PRB-77} Instead, other
experimental studies focused on the Raman spectra of bilayer
graphene. \cite{Malard-PRB,Yan-PRL-101,Das,Pimenta-phonon}
Recently, an experimental work on infrared spectra of gated
bilayer graphene as a function of doping appeared,
\cite{Kuzmenko-infrared} and a comparison with TB calculations
suggests that the TB prediction of the gate-induced band-gap is
overestimated.

In this work, we study by means of DFT ab initio calculations the
band-gap opening in bilayer graphene, both as a function of the
external electric field and as a function of doping. The paper is
organized as follows: in Sec. \ref{sec:theo} a description of the
system we investigate is reported, along with the computational
details. Results are presented in Sec. \ref{sec:res}, where the
dependence of the gap on the electric field and doping is first
shown and compared with TB calculations. Then, a detailed
analysis of the screening properties of the bilayer to the
external electric field is reported. The effect of the electronic
temperature on the screening is also investigated, and the
nonmonotonic behavior of the band-gap as a function of doping at
fixed electric field is explained. The $GW$ correction of the
DFT-calculated response of the gap to the external electric field
is presented. Finally, a general scheme to obtain the full band
structure of gated bilayer graphene is provided, and a comparison
with experimental findings is reported. In Sec. \ref{concl} our
conclusions are drawn. In the Appendix we describe in
detail the top and bottom gates implementation in our DFT
calculations.


\section{\label{sec:theo} Theoretical background}

\subsection{\label{general} Bilayer graphene in bottom and top gates}

\begin{figure}[t]
  \centering
   \includegraphics[width=1\columnwidth]{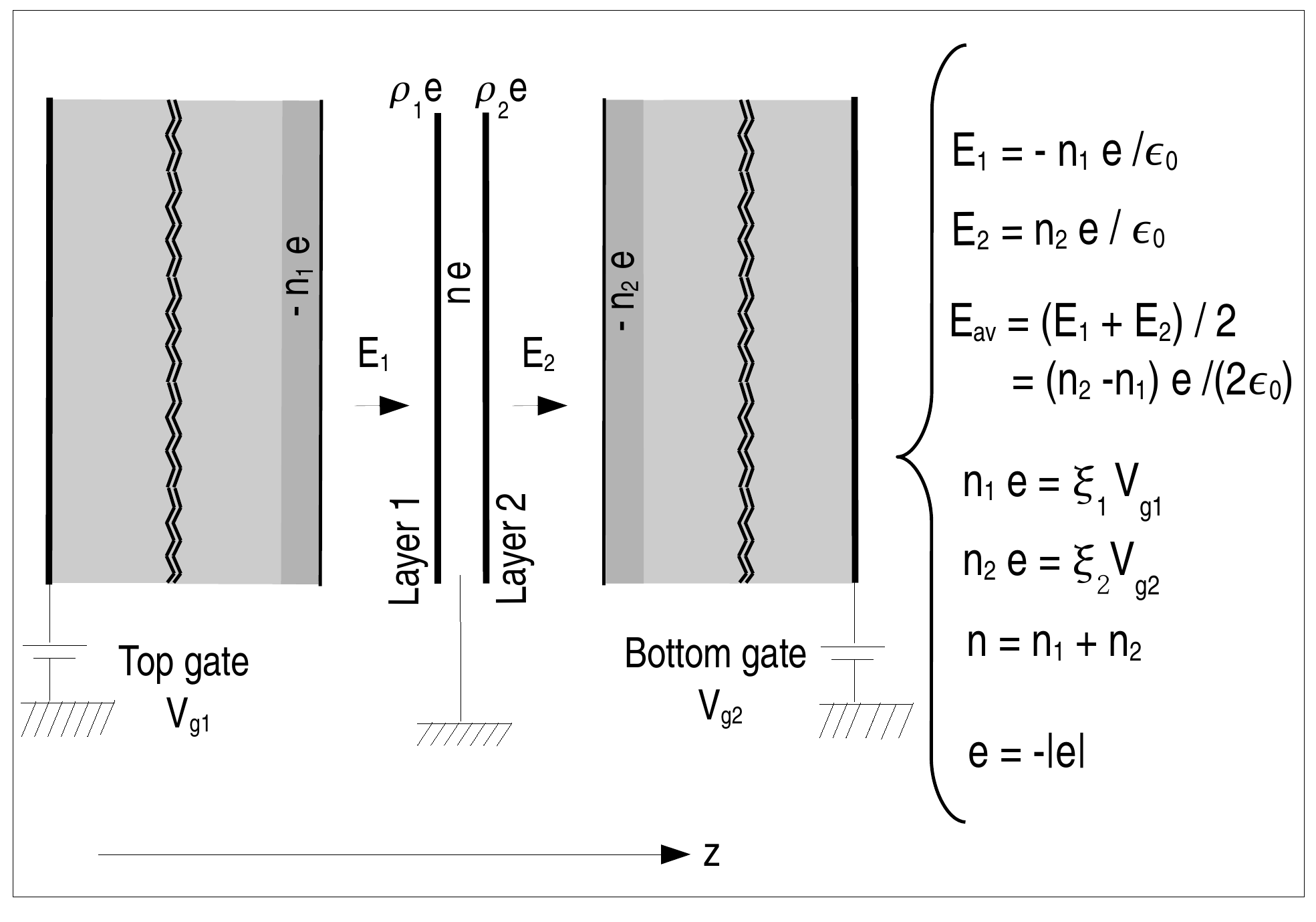} \\
  \caption{Schematic representation of the experimental setup where
           bilayer graphene is placed between two dielectric materials,
           and it is doped by applying bottom ($V_{g2}$) and top ($V_{g1}$) gates.
           The width of the two dielectrics is much larger than the distance between the dielectrics and the bilayer.
           A doping charge per unit area $ne = (n_1+ n_2)e$  is accumulated on the bilayer, where
           $n_2 e = \xi_2 V_{g2}$ and $n_1 e = \xi_1 V_{g1}$ are the charges from bottom
           and top gates, respectively.
           $\rho_1 e$ and $\rho_2 e$ are the electronic charges per unit area (with respect
           to the neutral case) on layer 1 and layer 2, respectively.
           In particular, $(\rho_1 + \rho_2)e=(n_1+ n_2)e$, while $\rho_1 \neq n_1$ and $\rho_2 \neq n_2$.
           Layer 1 and layer 2 of bilayer feel electric fields
           $E_1 = -n_1e/\epsilon_0$ and $E_2 = n_2e/\epsilon_0$, which determine an average
           electric field $E_{\rm{av}} = (n_2 - n_1)e/(2\epsilon_0)$.
           $\epsilon_0$ is the permittivity of the vacuum.
 }
  \label{setupexp}
\end{figure}

The experimental setup where bilayer graphene feels different bottom and
top gates is schematically represented in Fig.\ref{setupexp}.
The bilayer is first grown on a dielectric material, of width $D_2$ and relative dielectric constant $\epsilon_{r2}$.
Applying a voltage difference $V_{g2}$ (bottom gate) between the dielectric and the bilayer,
a doping charge per unit surface $n_2 e = \xi_2 V_{g2}$ is accumulated on the bilayer, where
 $e$ is the electron charge ($e$ = - $|e|$) and $\xi_2 = \epsilon_0 \epsilon_{r2}/D_2$.
$\epsilon_0$ is the permittivity of the vacuum.
Depositing another dielectric material of width $D_1$ and with relative dielectric constant $\epsilon_{r1}$
over the bilayer,
and applying a gate voltage $V_{g1}$ (top gate) between the dielectric and the bilayer,
 an additional doping charge per unit surface $n_1 e = \xi_1 V_{g1}$ is accumulated,
 where $\xi_1 = \epsilon_0 \epsilon_{r1}/D_1$.
A total doping charge $ne$ is therefore accumulated on the bilayer, where $n=n_1 + n_2$.
According to standard notation,
positive $n$ corresponds to electron doping and negative $n$ corresponds to
hole doping.
$\rho_1$ and $\rho_2$ are the electronic charges per unit area (with respect to the neutral case)
 accumulated on layer 1 and layer 2, respectively.
In particular, the sum of $\rho_1$ and $\rho_2$ is determined by the electrostatics, and
it is equal to the sum of $n_1$ and $n_2$. However, the individual values of $\rho_1$ and $\rho_2$
depend on the screening properties of the system, and in general $\rho_1 \neq n_1$ and $\rho_2 \neq n_2$.
In the configuration shown in Fig.\ref{setupexp}, layer 1 and layer 2 of the bilayer
 feel an electric field $E_1$ and $E_2$,
respectively, given by
\begin{eqnarray}
\label{E_1}
E_1 &=& -n_1\ e/\epsilon_0, \\
E_2 &=& n_2\ e/\epsilon_0.
\label{E_2}
\end{eqnarray}
The average electric field $E_{\rm{av}}$ is defined as
\begin{eqnarray}
E_{\rm{av}} &=& \left( E_1 + E_2 \right) / 2 \nonumber \\
&=&  (n_2 - n_1)\ e/(2\epsilon_0) \nonumber \\
&=&  (n_1 - n_2)\ |e|/(2\epsilon_0).
\label{Eav}
\end{eqnarray}
Positive $E_{\rm{av}}$ is oriented from dielectric 1 to dielectric 2 ($i.e.$, from top to bottom gate).
When $n_1$  and $n_2$ are equal, we are in the case
of equal bottom and top gates, and $E_{\rm{av}}$ vanishes. When $n_1$ is zero, we are in presence of
 bottom gate alone.
The top gate can also be generated by a chemical doping, with the deposition of alkali or halogen
atoms on the bilayer.
In this work, the electric fields $E_1$ and $E_2$ are simulated using periodically repeated
boundary conditions by introducing dipole and monopole potentials, as described in the Appendix.

\begin{figure}[t]
  \centering
  \includegraphics[width=1\columnwidth]{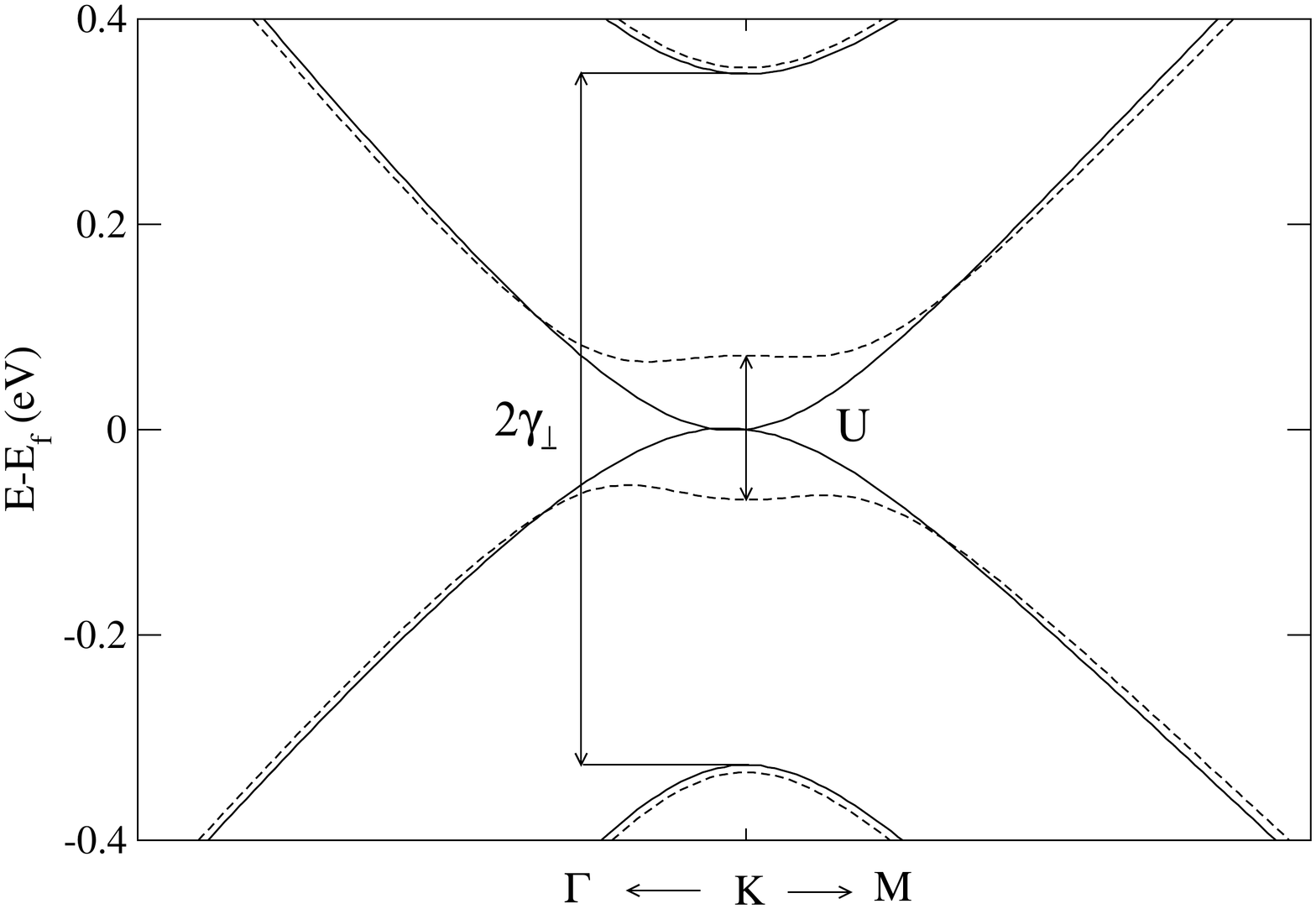} \\
  \caption{Band structure, around the $K$ point in the BZ,
           of undoped bilayer graphene in absence of bottom and
           top gates (solid line) and in presence of different bottom and top gates (dashed line).
 }
  \label{bande}
\end{figure}

The presence of different bottom and top gates generates an
electrostatic potential which is different on layer 1 with respect
to layer 2, and this asymmetry gives origin to a band-gap opening.
In Fig.\ref{bande} we show the band structure of undoped bilayer
graphene, in absence of bottom and top gates, where no gap is
observed, and in presence of different bottom and top gates in
which case a gap is opened. In order to simplify the discussion,
in this work we define a signed gap $U$ at the $K$ point in the
Brillouin zone (BZ), which is negative for $E_{\rm{av}} < 0$, and
positive for $E_{\rm{av}} > 0$.

\subsection{\label{sec:com} Computational details}

The presented ab initio results based on the
DFT, are done using
both the Perdew-Burke-Ernzerhof (PBE) (Ref.\cite{Perdew_PRL77})
GGA,
and the Perdew-Zunger (PZ) Ref.(\cite{PZ-LDA})
LDA exchange-correlation functionals.
Core electrons are taken into account using the pseudopotential method,
with norm-conserving Troullier-Martin pseudopotentials. \cite{TM-pseudo}
Plane-waves basis set is used to describe valence electron wave functions and density,
 up to a kinetic energy cut-off of 40 and 600~Ry, respectively.
The electronic eigenstates have been occupied with a Fermi-Dirac distribution,
using an electronic temperature of 300 and 30 K.
The BZ integration has been performed with a uniform $\mathbf{k}$ point grid of $(80\times 80 \times 1)$
and $(240\times 240 \times 1)$ for the two temperatures, respectively.
The experimental lattice constant $a$ = 2.46 \cf{\AA} of two-dimensional graphite is used.
The layer-layer distance $d$ is fixed at the value of 3.35 \cf{\AA}, as in graphite.
The two layers are arranged according to Bernal stacking.
The length $L$ of the supercell along $z$ is 17.2 \cf{\AA}.
Calculations have been performed using the {\tt PWscf} code \cite{PWSCF_WEB} of the {\tt Quantum ESPRESSO}
distribution \cite{ESPRESSO_WEB}.

\begin{figure}[t]
  \centering
  \includegraphics[width=0.8\columnwidth]{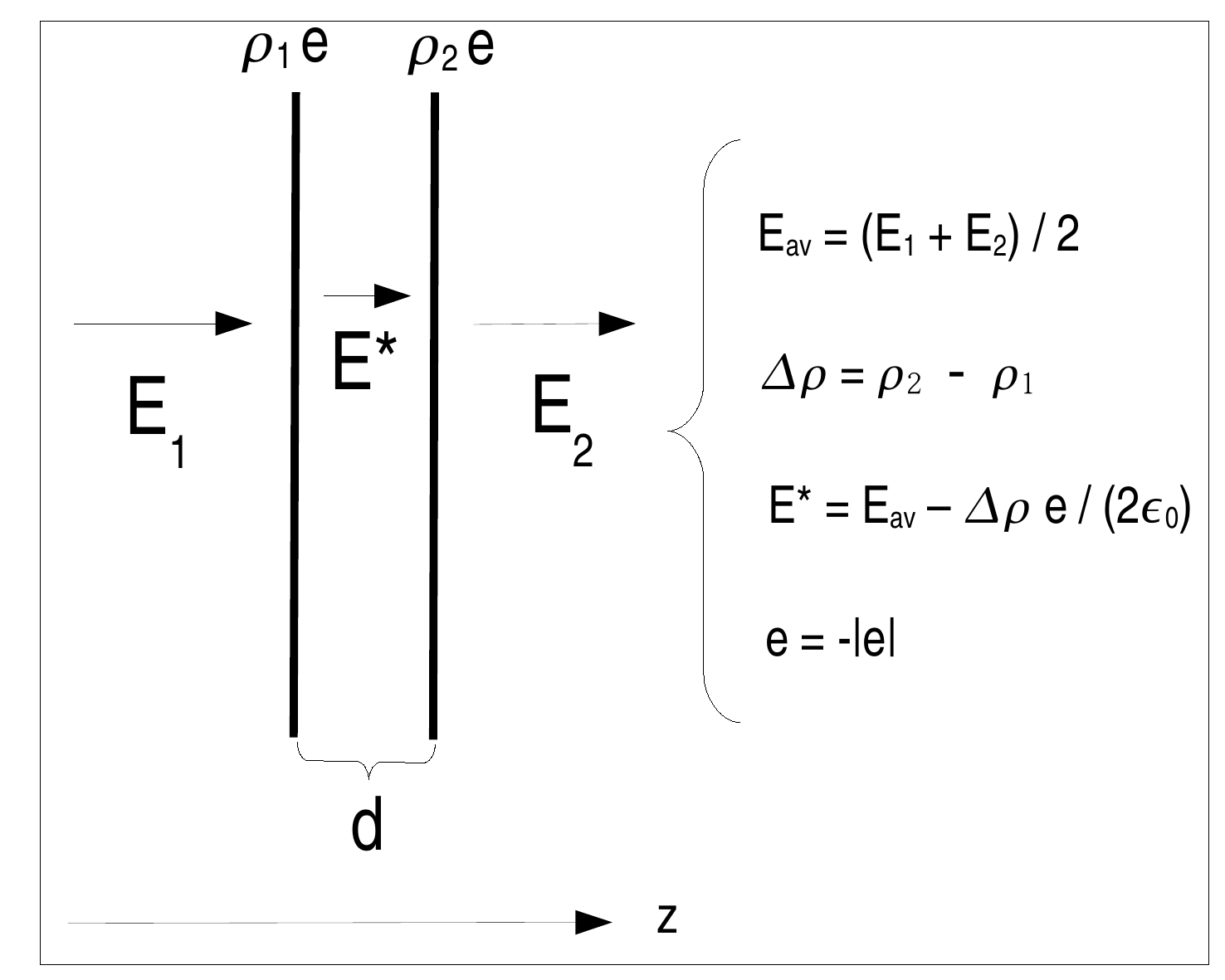}
  \caption{ Schematic representation of the electrostatic model used with TB calculations. The
           bilayer of thickness $d$ feels an external electric field
           $E_{1}$ on layer 1 and $E_{2}$ on layer 2, which induce a charge
           per unit surface $\rho_1 e$ and $\rho_2 e$
           on layer 1 and layer 2 and a local electric field $E^*$.
 }
  \label{fig:dielectric}
\end{figure}

In this work we perform also TB calculations.
The TB model we use is characterized by two parameters, $\gamma_{\parallel}$ and $\gamma_{\perp}$,
which represent the first nearest-neighbors in-plane hopping and the interplane hopping
between vertically superposed atoms in the Bernal stacking configuration, respectively.
$\gamma_{\parallel}$ is related to the Fermi velocity in single layer graphene,
$v_f = \gamma_{\parallel} a \sqrt{3}/(2\hbar)$.
We use a value of $\gamma_{\parallel}$= 3.1~eV,
as inferred from experimental measurements.~\cite{Ohta_science,Malard-PRB,Novoselov}
Within the TB model, $2 \gamma_{\perp}$ corresponds
to the band splitting between the lowest occupied $\pi$ band and the highest unoccupied $\pi$ band at $K$
(see Fig.\ref{bande}).
We use a value of $ \gamma_{\perp}$ = 0.4~eV, as obtained
from experimental measurements.~\cite{Ohta_science,Malard-PRB,Kuzmenko-infrared}
Similar values for these TB parameters have been used in literature. \cite{McCann-PRB-74,Castroneto_PRL,Castroneto_arXiv}
In addition to these parameters, we consider the energy difference between layer 2 and layer 1
induced by the electric field, which coincides with the signed gap $U$ at the $K$ point in the BZ
(see Fig.\ref{bande} and Ref.\cite{Ohta_science}).

In the TB formalism, in order to obtain the relation between the gap $U$ and the average electric field $E_{\rm{av}}$,
a simple electrostatic model is used.
In Fig.\ref{fig:dielectric} we show a schematic representation of this electrostatic model.
Charges per unit surfaces $\rho_1e$ and $\rho_2e$ are concentrated on the two layers
of the bilayer, which create a screened field $E^*$ inside the system.
Using simple electrostatic equations, we have
\begin{equation}
E^* = E_{\rm{av}} -\ \frac{\Delta \rho\ e}{2\epsilon_0},
\label{Estar}
\end{equation}
where $E_{\rm{av}}= (E_1 + E_2)/2$ and $\Delta \rho = \rho_2 - \rho_1$.
$\Delta \rho $ is calculated from the square modulus of the eigenfunctions
in the two layers.
The energy difference between layer 2 and layer 1, $i.e.$, the band-gap $U$ at $K$, is given by
\begin{equation}
U =\ -\ d\ E^*\ e.
\label{deltaV}
\end{equation}
Inserting into eq.(\ref{deltaV}) the expression of $E^*$ as given in eq.(\ref{Estar}),
and writing $E_1$ and $E_2$ as in Eq.(\ref{E_1}) and (\ref{E_2}), we obtain:
\begin{equation}
U  =\ \frac{d e^2}{2\epsilon_0}\  (n_1 - n_2 + \Delta \rho).
\label{eq-for-TB}
\end{equation}
Therefore, in the TB calculations the electronic screening is evaluated using
the simplified electrostatic model described above, contrary to the DFT formalism where
the detailed shape of the charge distribution is fully taken into account.


\section{\label{sec:res} Results}


\subsection{\label{sec:alpharho-U} Band gap as a function of the external electric field and doping charge}

As anticipated in Sec.\ref{general}, when bilayer graphene feels
different bottom and top gates, a band-gap $U$ is opened. In this
section we first investigate the dependence of $U$ on the average
external electric field $E_{\rm{av}}$, at fixed doping $n$.

\begin{figure}[t]
  \centering
  \includegraphics[width=1.1\columnwidth]{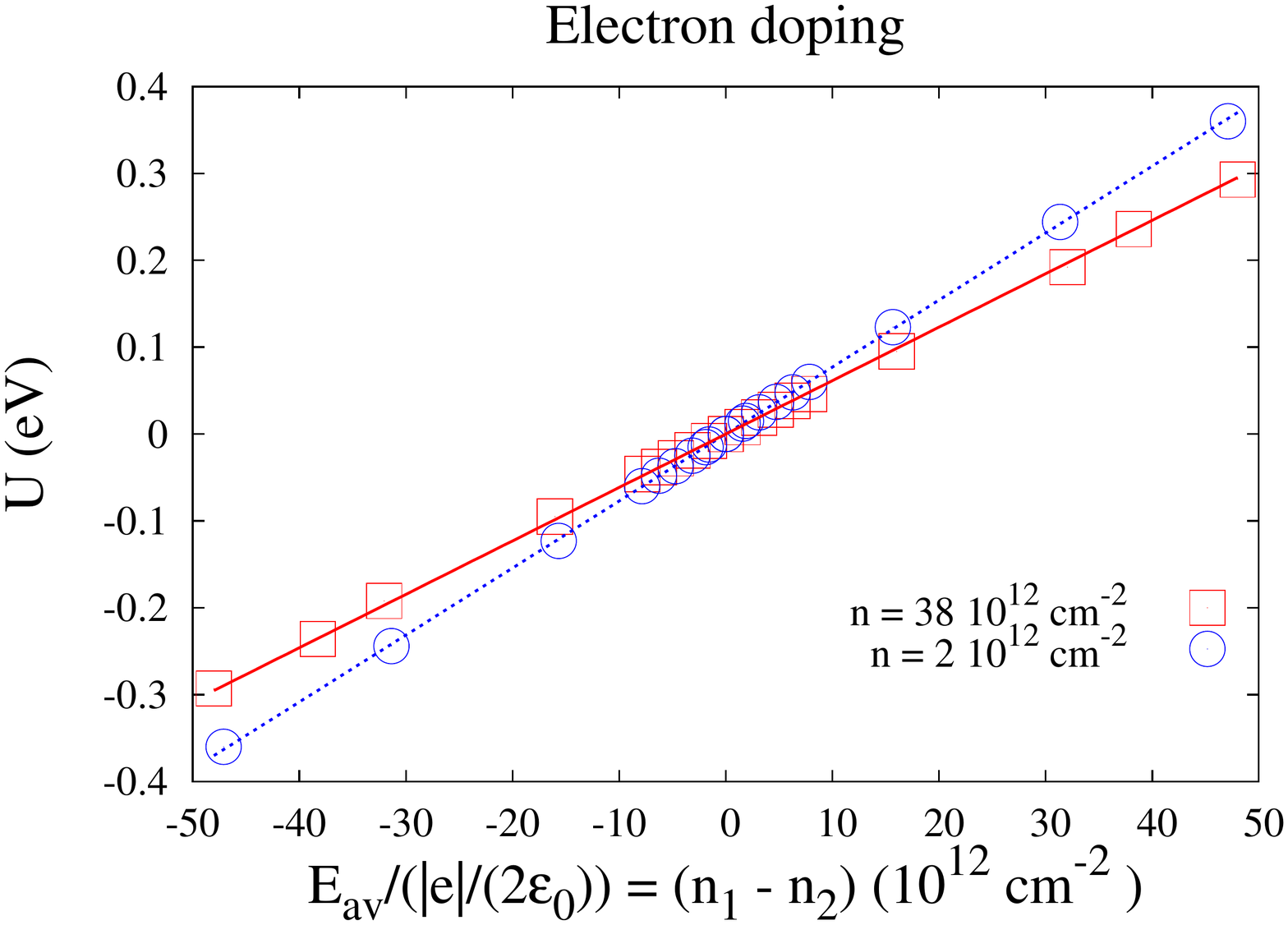} \\
  \includegraphics[width=1.1\columnwidth]{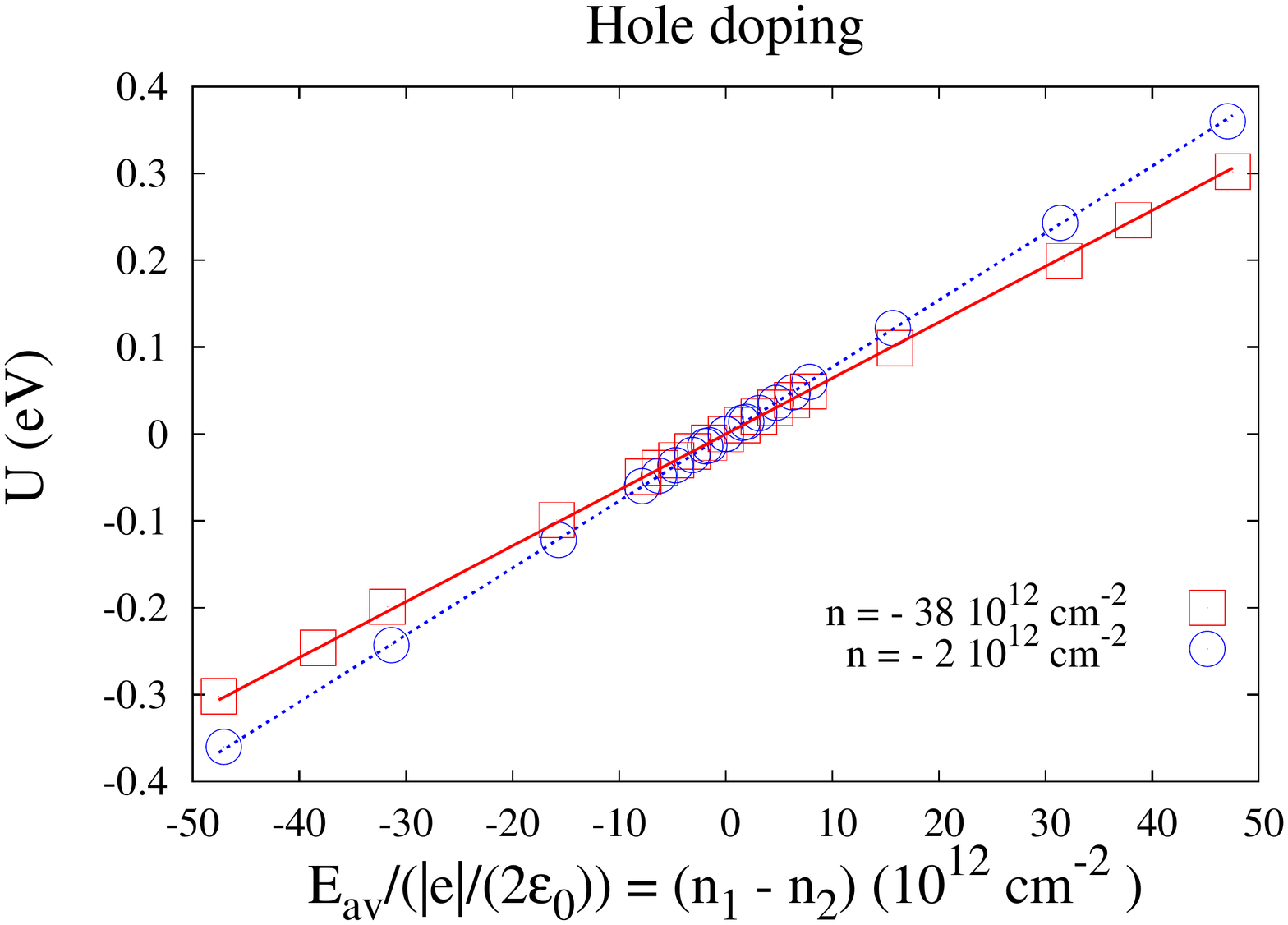}
  \caption{ (Color online) DFT-GGA calculated $U$ as a function of $(n_1 - n_2)$,
            $i.e.$, the average electric field divided by $|e|/(2\epsilon_0)$,
            for electron and hole dopings.
             Two values of electron doping are shown,
            $n$ = 38 and 2 \cf{\times\ 10^{12}\ cm^{-2}}, and two values of hole doping,
            $n$ = -38 and -2 \cf{\times\ 10^{12}\  cm^{-2}}. Points are the calculated values,
            for an electronic temperature of 300 K, while lines are linear fits.
 }
  \label{fig:bandgap-e-h}
\end{figure}

In Fig.\ref{fig:bandgap-e-h} we show the DFT-GGA calculated $U$ as
a function of $(n_1 - n_2)$ [$i.e.$, the average electric field
$E_{\rm{av}}$ divided by $|e|/(2\epsilon_0)$], for two values of
electron and hole dopings. These values of doping are chosen as
representative of two different doping regimes, which can be
experimentally obtained in bilayer graphene by the application of
a gate voltage with a \cf{SiO_2} dielectric \cite{Yan-PRL-101} or
with a polymeric electrolyte. \cite{Das} Our results show that $U$
has a linear dependence on the applied electric field
$E_{\rm{av}}$. We therefore define a linear response $\alpha (n)$
such that
\begin{equation}
U\ (n,E_{\rm{av}})= \alpha (n) (n_1 - n_2).
\label{def-alphagap}
\end{equation}

\begin{figure}[t]
  \centering
  \includegraphics[width=1.1\columnwidth]{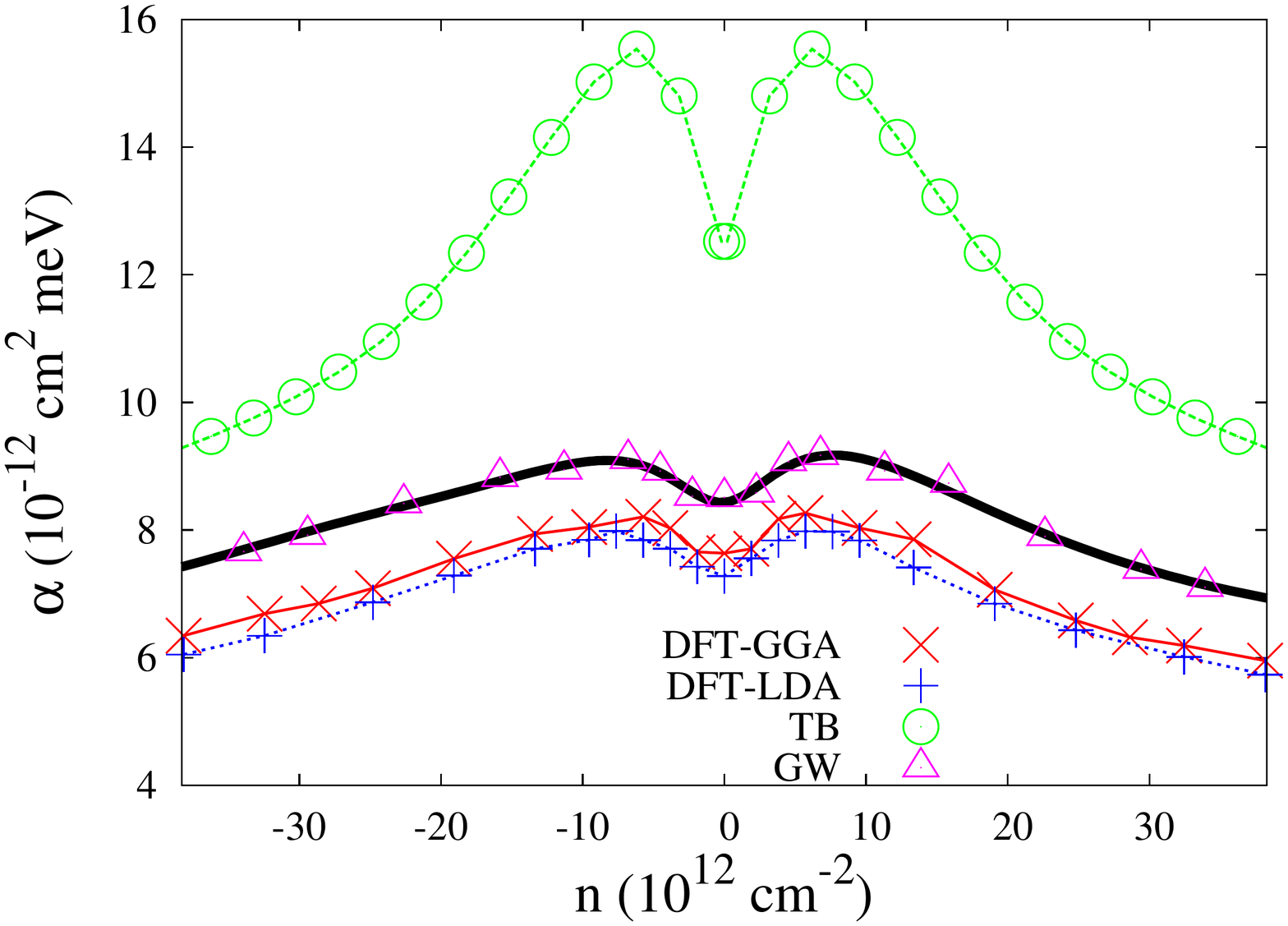} \\
  \caption{ (Color online) $\alpha$ as a function of doping $n$, for an electronic temperature of
   300 K, calculated with DFT-GGA ($\times$ crosses) and LDA ($+$ crosses), calculated using TB model
   with $\gamma_{\parallel}$ = 3.1  eV and $\gamma_{\perp}$ = 0.4 eV (circles),
   and using $GW$ correction (up-triangles).
   The $GW$ correction is obtained as described in
   Sec. \ref{GW}.
   The continuous thick (black) line is the fit,
   as written in Eq.(\ref{fit-lorent}), of the $GW$ result.
   The value of $\alpha$ in absence of electronic screening is
   independent on the doping, and it is $\alpha^{\cf{bare}}=d e^2/(2\epsilon_0)$ = 30.3 \cf{\times\ 10^{-12}\ cm^{2}\ meV}.
 }
  \label{fig:alpha-gap}
\end{figure}

In Fig.\ref{fig:alpha-gap} we show $\alpha$ as a function of doping $n$,
calculated from Eq.(\ref{def-alphagap}),
for an electronic temperature of 300 K, within the DFT-GGA and LDA functionals, and
using the TB model described in Sec.\ref{sec:com}.
Contrary to previous results in literature, \cite{Min-PRB-75,Aoki-solstatecomm-142}
our LDA and GGA results are very similar, and not in agreement with the TB ones.
In particular, our results for zero doping and GGA functional are in agreement with the previous GGA study.
\cite{Min-PRB-75} They disagree instead with the ones computed with LDA
functional in Ref.\cite{Aoki-solstatecomm-142}.
This is probably due to the fact that in Ref.\cite{Aoki-solstatecomm-142} the authors used a
coarse $\mathbf{k}$ point sampling (10x10x1) with respect to the ones used
in this work and in Ref.\cite{Min-PRB-75}, and
their results are likely unconverged.
In the following we only present our GGA results.
Both DFT and TB $\alpha$'s display a nonmonotonic
behavior as a function of doping.
However, the values of the DFT-calculated $\alpha$
are substantially different from the TB ones, especially for low doping values,
$i.e.$, when the Fermi level is close to the band-gap edges.
This is the most interesting case for the application of the bilayer as active device in electronics.

We notice that in the absence of electronic screening, $\alpha$ is independent
of the doping, and it is
\begin{equation}
\alpha^{\cf{bare}} = \frac{d e^2}{2\epsilon_0} \ =\ 30.3 \times 10^{-12} \cf{cm^{2}\ meV}.
\end{equation}
Thus, with the inclusion of
the electronic screening, the DFT-calculated $\alpha$ becomes roughly
three times smaller than the $\alpha^{\cf{bare}}$,
which suggests that the screening effects are crucial for the description of the band-gap.

In order to understand the origin of the difference between the DFT and TB results,
we notice that this can be due $(i)$ to the calculated electronic band
structure and charge transfer and
$(ii)$ to the electrostatic model used in the TB
calculations, which gives a simplified description of the crucial screening effects, fully
included in the DFT formalism.
To verify the quality of the electrostatic model, we introduce
the quantity $\eta (n)$ defined as:
\begin{equation}
\Delta \rho\ (n,U)=\ \eta (n) U.
\label{def-eta}
\end{equation}
$\Delta \rho = \rho_2 - \rho_1$ is calculated from
\begin{eqnarray}
\rho_2 = \int^{\infty}_{0} \left[\rho (z)-\rho_0 (z)\right]\ dz, \\
\rho_1 =  \int^{0}_{-\infty} \left[ \rho (z)-\rho_0 (z)\right]\ dz,
\end{eqnarray}
where $\rho (z)$ is the planar average of the electronic charge density (per unit volume)
for a doping $n$ and in presence of $E_{\rm{av}}$ and $\rho_0 (z)$
is the planar average of the electronic charge density (per unit volume)
for the neutral case, with $E_{\rm{av}}$=0.
Here and in the following $z=0$ indicates the plane at the midpoint of
the two graphene layers.
In our DFT calculations, $\pm \infty$ corresponds to $\pm L/2$, where
$L$ is the length of the supercell along $z$.

$\eta$ is a measure of the charge transfer between layers
in the presence of a band-gap $U$.
We introduce this quantity because it is a direct outcome of the TB calculations,
and no further electrostatic model is needed to compute it.
Moreover, according to the electrostatic model used together with the TB formalism,
described in Sec. \ref{sec:com},
the relation which gives $\alpha$ as a function of $\eta$
is obtained dividing Eq.(\ref{eq-for-TB}) by U,
\begin{equation}
\alpha (n)= \frac{\alpha^{\cf{bare}}}{1\ -\  \eta (n)\ \alpha^{\cf{bare}}}.
\label{formuletta-wrong}
\end{equation}

\begin{figure}[t]
  \centering
  \includegraphics[width=1.1\columnwidth]{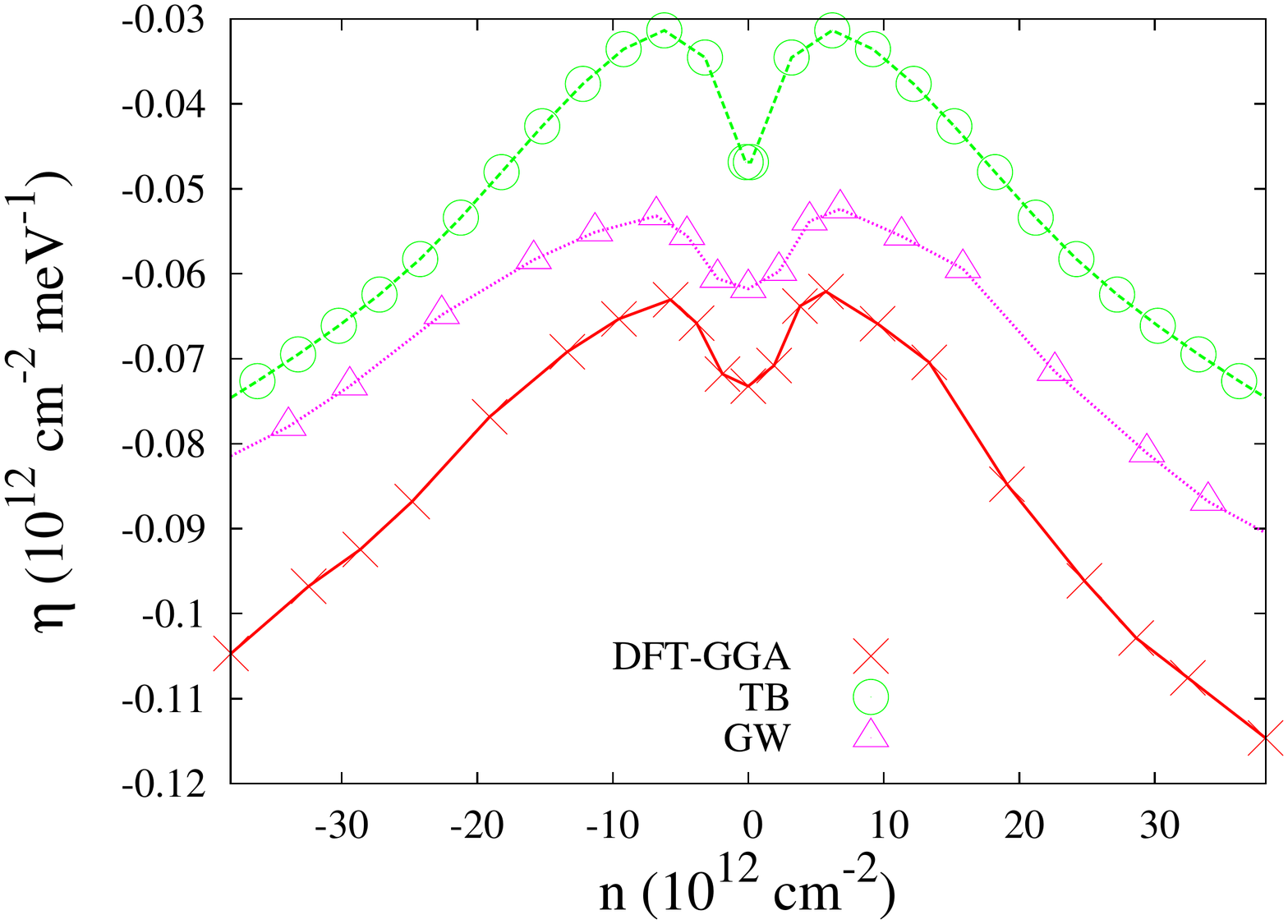}
  \caption{ (Color online) $\eta$ as a function of doping $n$, calculated with DFT-GGA
           (crosses), with TB model (circles), and with $GW$ correction (up-triangles),
           for an electronic temperature of 300 K.
           The $GW$ correction is obtained as described in
   Sec. \ref{GW}.
 }
  \label{fig:etaabinit}
\end{figure}

In Fig.\ref{fig:etaabinit} we show $\eta(n)$, calculated from Eq.(\ref{def-eta}),
for an electronic temperature of 300 K, within the DFT and
using the TB model described in Sec.\ref{sec:com}.
$\eta$ as a function of $n$ has a nonmonotonic behavior as found for $\alpha(n)$,
and this trend is well described by the two methods.
However, the values of $\eta$ calculated with the two formalisms are different.
Since no electrostatic model is used in the TB calculations, we conclude that this discrepancy
originates only from the difference between the DFT- and TB-calculated band structure and charge transfer.

Moreover, comparing the DFT and TB results of $\alpha(n)$ and $\eta(n)$,
we see that, for low doping levels, the relative difference, with respect to DFT values,
of the TB/DFT $\alpha(n)$'s is around 60\%, while the analogous difference for $\eta$ is around 30\%.
Therefore, the electrostatic model used to compute $\alpha (n)$ in the TB formalism
introduces a large error in the description of the band-gap opening in presence of
an external electric field.


\subsection{\label{sec:screening}Electronic screening effects}

In this section we analyze where the simplified electrostatic model used in the TB
calculations fails, and we propose a more sophisticated one.
\begin{figure}[t]
  \centering
  \includegraphics[width=1.1\columnwidth]{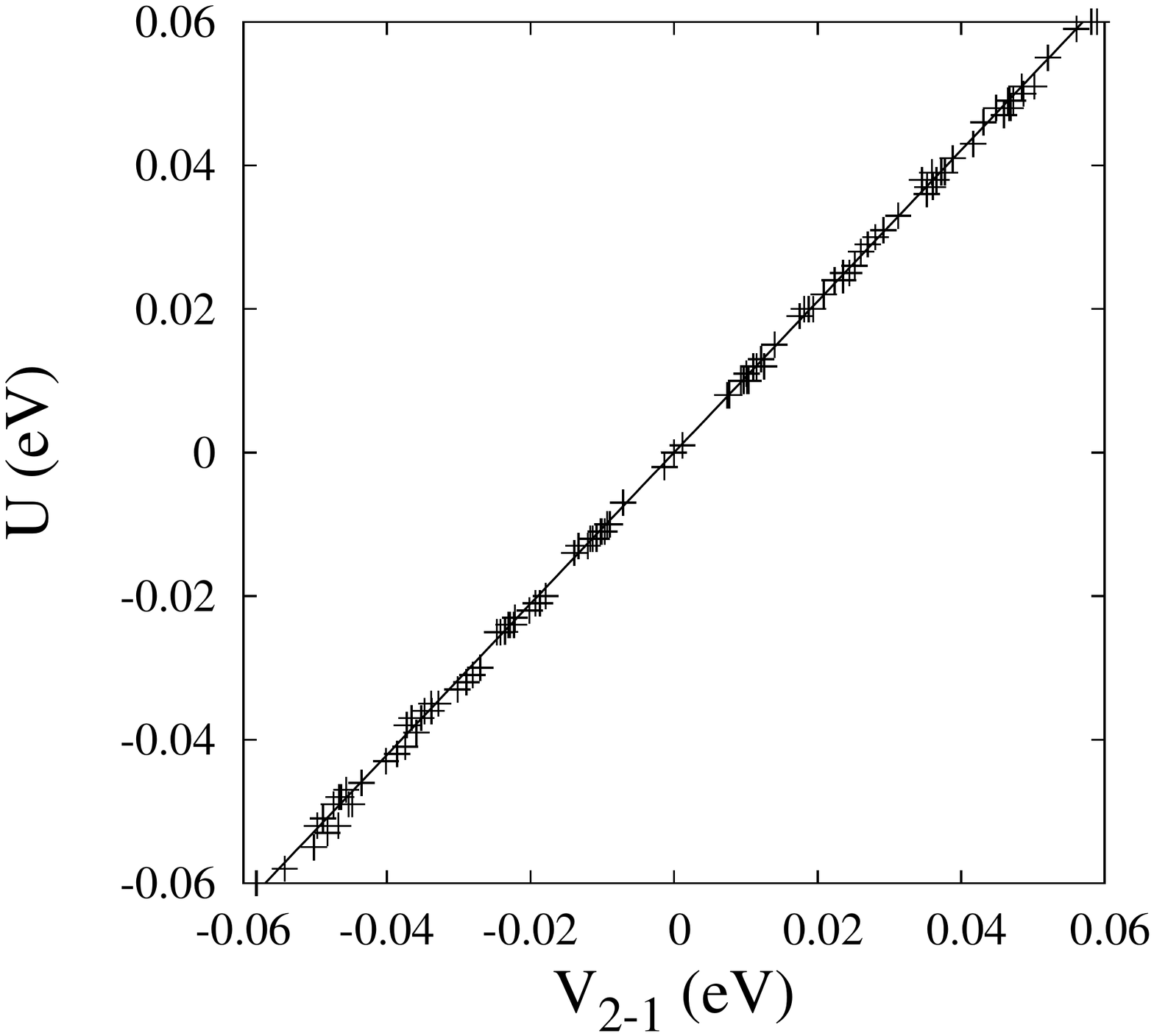}
  \caption{DFT-GGA calculated $U$ as a function of the difference between the planar average of the
           ionic, Hartree, and electrostatic potential energy on layer 2 and layer 1,
           $V_{2-1}$ , for different values of doping and of $E_{\rm{av}}$.
           Points are the calculated values; the line is the linear fit, $U$ = 1.072 $\times\ V_{2-1}$.
           }
  \label{fig:plot-deltaV}
\end{figure}
First of all, we know that in the TB formalism the energy difference between the two layers
coincides with the band-gap $U$ at the $K$ point in the BZ.
In Fig.\ref{fig:plot-deltaV} we show the band-gap $U$ as a function of $V_{2-1}= V_2 -V_1$,
where $V_2$ and $V_1$ are the planar average of the DFT-calculated ionic, Hartree, and electrostatic potential
energy on layer 2 and layer 1, respectively.
The inclusion of the exchange-correlation potential does not change the result.
We can notice that even in DFT formalism, $U$ is correlated with the potential energy difference between
the two layers, and in particular,
\begin{equation}
U\ =\  \beta V_{2-1},
\label{beta}
\end{equation}
where $\beta = 1.072$,
slightly higher than the expected unitary slope.

To better understand the screening effects in the system,
we investigate the linearly induced charge (per unit volume) $\rho^{(1)}$
\begin{eqnarray}
\label{prima}
&\rho^{(1)}& (z;n,E_{\rm{av}})=\frac{\partial{\rho (z;n,E_{\rm{av}})}}{\partial{E_{\rm{av}}}}\  E_{\rm{av}} \\
                        &\simeq & \frac{1}{2} \left[\rho (z;n,E_{\rm{a}v})-\rho (z;n,-E_{\rm{av}})\right],
\label{seconda}
\end{eqnarray}
where $\rho (z;n,E_{\rm{av}})$ is the planar average of the charge density (per unit volume) at a given doping level $n$
and in presence of an external average electric field $E_{\rm{av}}$.
Such $\rho^{(1)}$ is antisymmetric with respect to $z$=0, $i.e.$, $\rho^{(1)} (z;n,E_{\rm{av}})= - \rho^{(1)} (-z;n,E_{\rm{av}})$.
In our plots we use the finite difference expression of $\rho^{(1)}$, $i.e.$, Eq.(\ref{seconda}).

\begin{figure}[t]
  \centering
  \includegraphics[width=1.04\columnwidth]{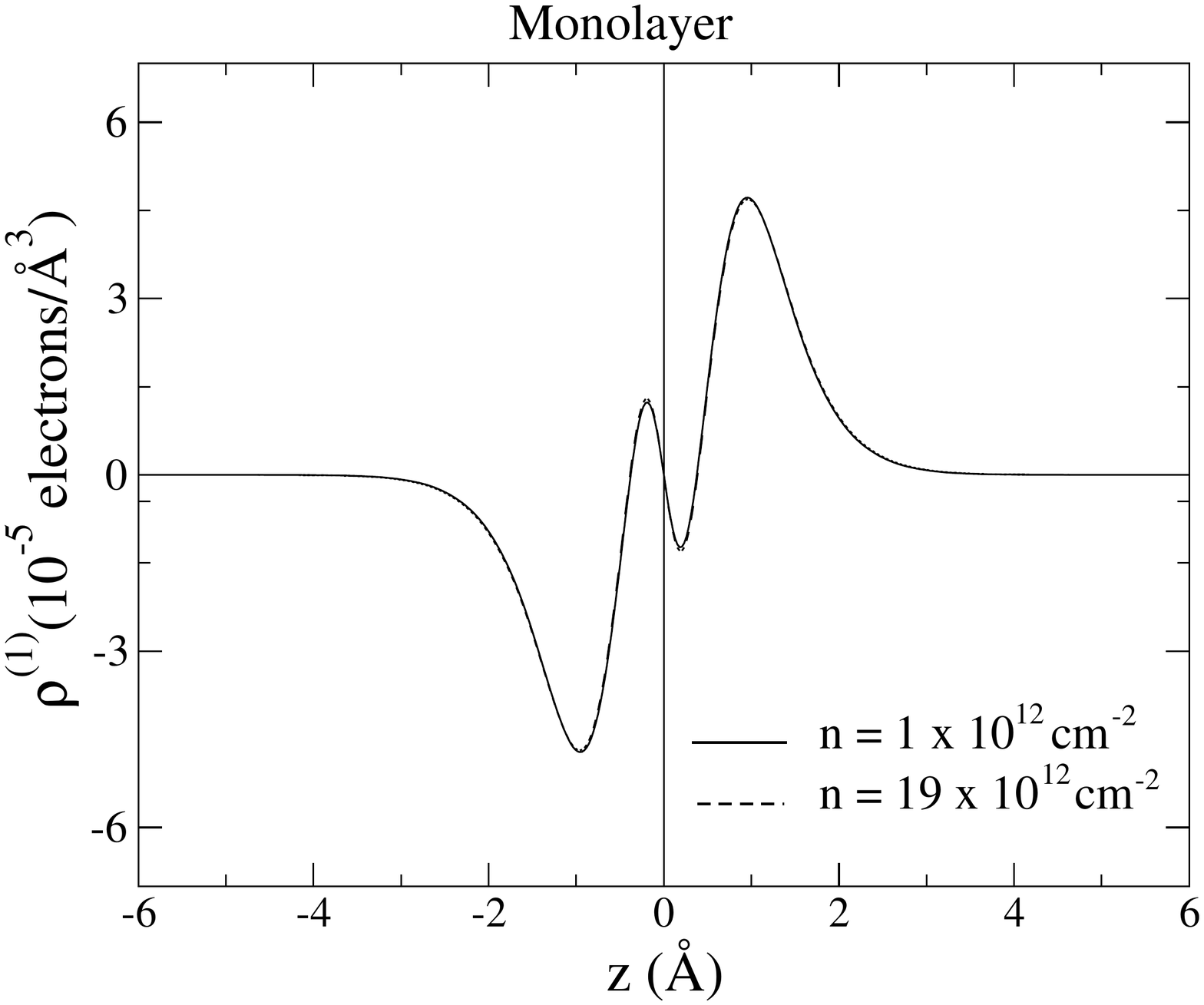}
  \caption{ Planar average of the linear induced charge (per unit volume)
            $\rho^{(1)}$ of a graphene monolayer in presence of
           an external electric field $E_{\rm{av}} = 1.6 \times e/(2\epsilon_0)$ \cf{10^{12} cm^{-2}}
           for a doping level $n$ = 1 \cf{\times\ 10^{12} cm^{-2}} (continuous line) and
           $n$ = 19 \cf{\times\ 10^{12} cm^{-2}} (dashed line).
 }
  \label{charge-density-mono}
\end{figure}

In Fig.\ref{charge-density-mono} we show $\rho^{(1)}$ for the graphene monolayer
in presence of an external electric field $E_{\rm{av}} = 1.6 \times e/(2\epsilon_0)$ \cf{10^{12} cm^{-2}} for
two different doping levels.
In this case, obviously no charge transfer between layers occurs, and the
electronic screening to the external electric field is only characterized
by an intralayer polarization. Moreover, we notice that the dependence of the induced charge
on the doping is negligible.

\begin{figure}[!t]
  \centering
  \begin{tabular}[t]{c}
   \includegraphics[width=1\columnwidth]{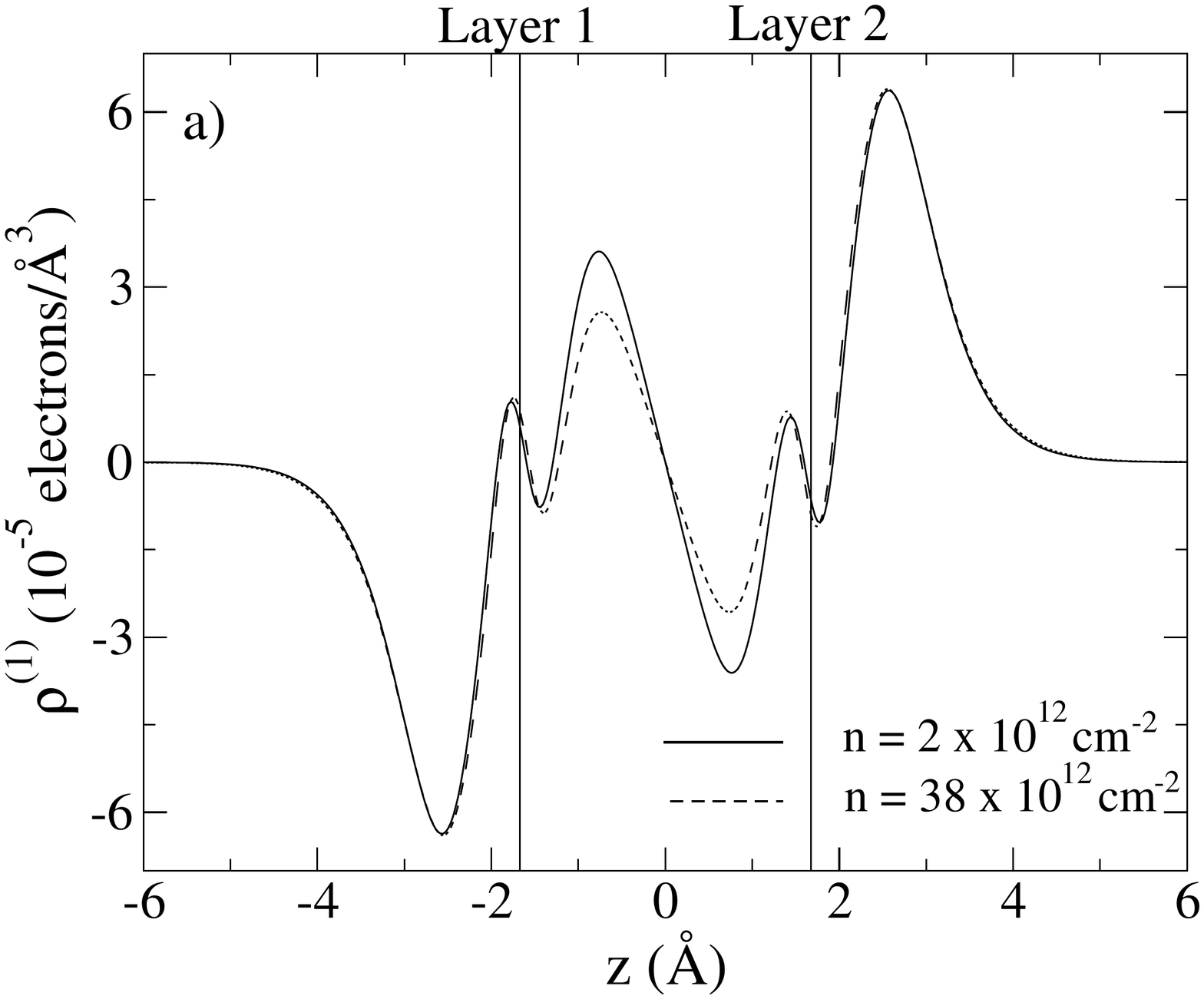} \\
   \includegraphics[width=1\columnwidth]{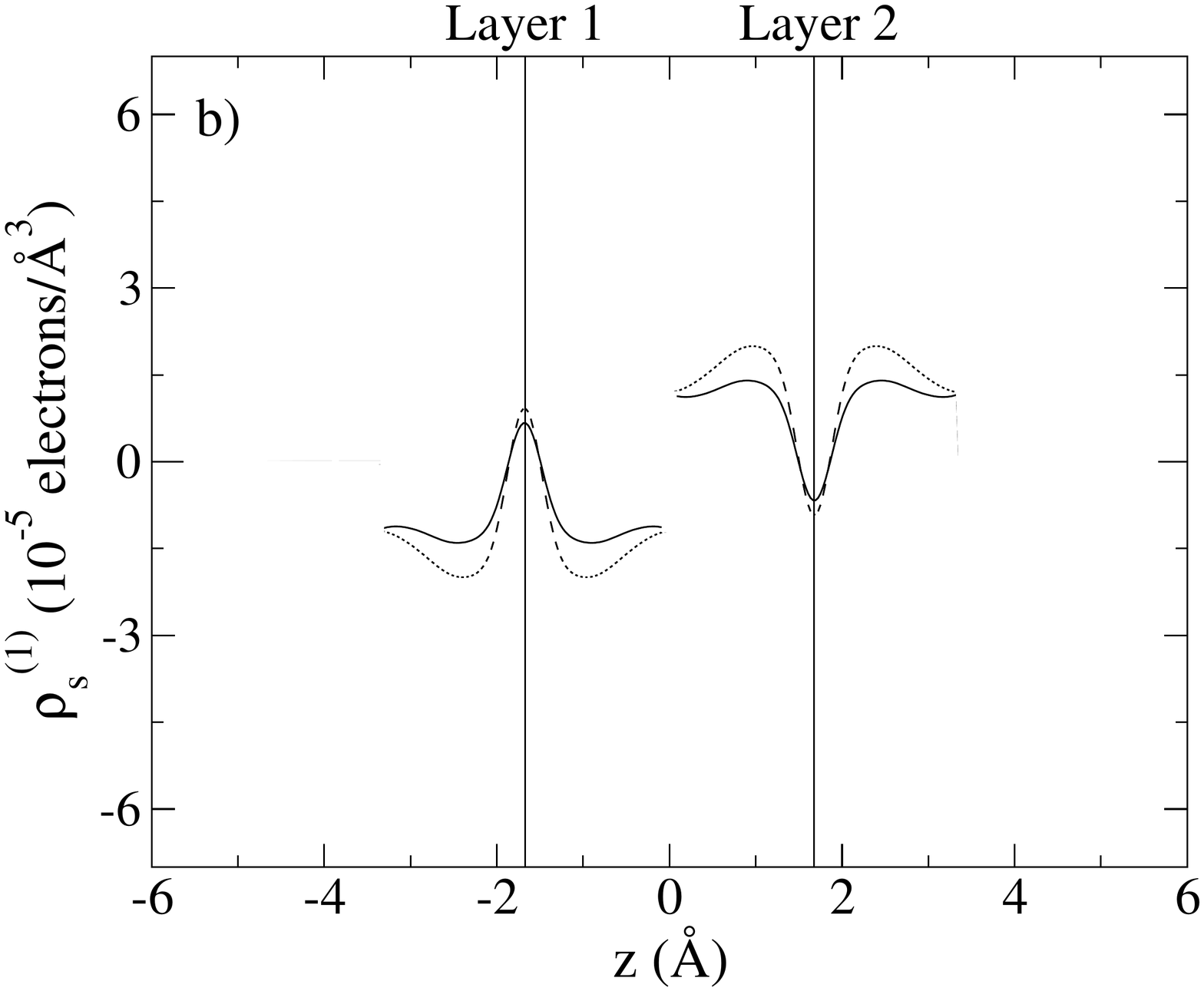} \\
   \includegraphics[width=1\columnwidth]{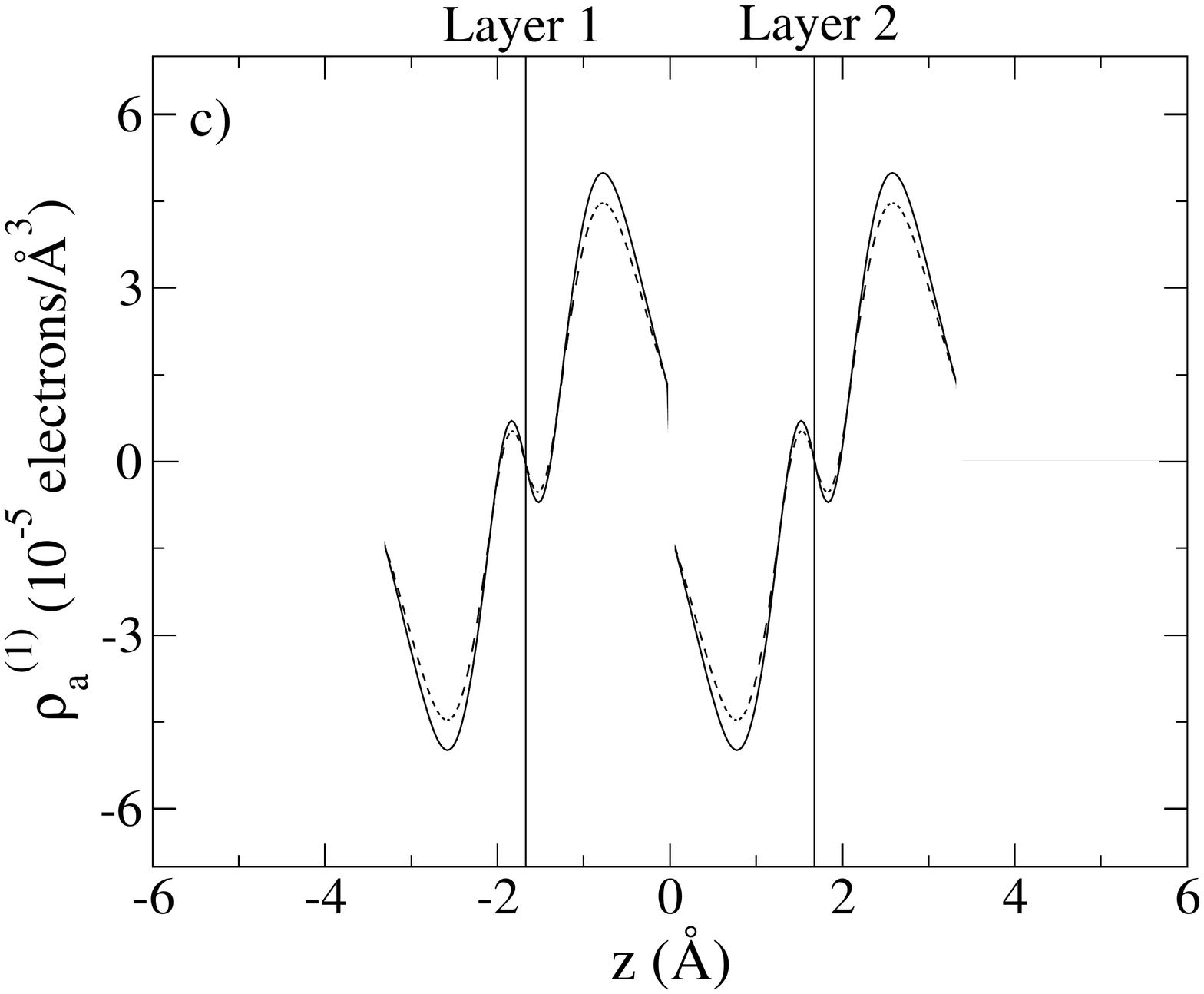} \\
  \end{tabular}
  \caption{ a) Planar average of the linearly induced charge (per unit volume) $\rho^{(1)}$ for bilayer graphene
           in presence of an external electric field $E_{\rm{av}} = 1.6 \times e/(2\epsilon_0)$  \cf{10^{12} cm^{-2}}
           for a doping level $n$ = 2 \cf{\times\ 10^{12} cm^{-2}} (continuous line) and
           $n$ = 38 \cf{\times\ 10^{12} cm^{-2}} (dashed line);
           b) symmetric component, $\rho^{(1)}_{s}$; and c) antisymmetric component, $\rho^{(1)}_{a}$,
           with respect to each layer, of the linearly induced charge $\rho^{(1)}$ shown in a)
           for the same doping levels.
 }
  \label{charge-density}
\end{figure}

In Fig.\ref{charge-density}-a) we show $\rho^{(1)}$ for bilayer
graphene in presence of the same  external electric field. First
of all, we notice that $\rho^{(1)}$ in the monolayer and in the
bilayer are of the same order of magnitude. Then, we observe that
the electronic screening of the bilayer to the external electric
field, is characterized by $(i)$ a charge transfer between the two
layers, which is peculiar to the bilayer and $(ii)$ an intralayer
polarization, which is also present in the monolayer.

In order to separate in the bilayer the interlayer from the
intralayer polarization, we notice from Fig.\ref{charge-density-mono}
that the intralayer induced charge is antisymmetric with respect to each individual layer.
Thus we decompose the induced charge in the bilayer into a symmetric component, $\rho^{(1)}_{s}$,
and an antisymmetric component, $\rho^{(1)}_{a}$, with respect to each individual layer.
$\rho^{(1)}_{s}$ and $\rho^{(1)}_{a}$ are defined for $z\in\{-d;d\}$,
$i.e.$, in an interval of width $d$ around each layer, where $d$ is the intralayer distance;
they are calculated as

\begin{equation}
\rho^{(1)}_{s/a}(z) = \frac{1}{2}\left\{\rho^{(1)}(z) \pm \rho^{(1)}\left[\cf{sign}(z)\ d-z\right]\right\}.
\label{zmin0}
\end{equation}
The symmetric, $\rho^{(1)}_{s}$, and antisymmetric, $\rho^{(1)}_{a}$, components
are related to the charge transfer between the two layers and to the intralayer polarization,
respectively.

In Fig.\ref{charge-density}-b) we show the symmetric component $\rho^{(1)}_{s}$, with respect to each layer,
of the induced charge $\rho^{(1)}$ shown in Fig.\ref{charge-density}-a).
In Fig.\ref{charge-density}-c) we show the antisymmetric component $\rho^{(1)}_{a}$.
In particular, $\rho^{(1)}_{a}$ is very similar
to the induced charge in the monolayer (Fig.\ref{charge-density-mono}),
and it is of the same order of magnitude of the total induced charge in the bilayer [Fig.\ref{charge-density}-a].
On the basis of this qualitative analysis of the linearly induced charge,
 we conclude that the intralayer polarization, which is not taken into account
in the TB formalism, gives an important contribution to the screening properties
of the system.

In order to quantify the effect of the induced charge on the gap,
we write the exact expression of
the potential energy difference $V_{2-1}$ in terms of the
linearly induced charge $\rho^{(1)}$ and of the external average electric field $E_{\rm{av}}$
using the Poisson equation in one dimension.
We obtain the following:
\begin{eqnarray}
V_{2-1}&=&V(d/2)\ -\ V(-d/2)\ = \ - \ de\ E_{\rm{av}}\ + \nonumber \\
&-& \frac{\displaystyle e^2}{\displaystyle 2\epsilon_0} \int^{+\infty}_{-\infty} |\frac{d}{2}-z|\  \rho^{(1)} (z)\ dz +\nonumber  \\
&+& \frac{\displaystyle e^2}{\displaystyle 2\epsilon_0} \int^{+\infty}_{-\infty} |-\frac{d}{2}-z|\ \rho^{(1)} (z)\ dz,
\end{eqnarray}
where $\pm\ d/2=\pm$ 1.675\ \cf{\AA} is the $z$ coordinate of the two layers.
Considering that $\rho^{(1)} (z) = -\rho^{(1)} (-z)$, by simple algebra and without approximations
we can rewrite $V_{2-1}$ as
\begin{equation}
V_{2-1}\ = \ -\ ed\ E_{\rm{av}}\ +\ \frac{\displaystyle d e^2}{\displaystyle 2\epsilon_0}\ \Delta \rho\  +\ e\ D_a\ - \  e\ D_s,
\label{nuova-formuletta}
\end{equation}
where
\begin{eqnarray}
\label{DA}
D_a\ =\ \frac{\displaystyle e}{\displaystyle \epsilon_0}\int^{d}_{0}\ (z-\frac{d}{2})\ \rho^{(1)}_a (z)\ dz, \\
D_s\ =\ \frac{\displaystyle e}{\displaystyle \epsilon_0}\int^{d}_{0}\ |z-\frac{d}{2}|\ \rho^{(1)}_s (z)\ dz,
\label{DS}
\end{eqnarray}
and
\begin{equation}
\Delta \rho = \int^{\infty}_{0} \rho^{(1)} (z)\ dz\ -\ \int^{0}_{-\infty} \rho^{(1)} (z)\ dz .
\end{equation}

$D_a$ and $D_s$ represent the contributions to the potential energy difference $V_{2-1}$
given by the antisymmetric and symmetric components of the linearly induced charge around
each layer.
Equation (\ref{nuova-formuletta}) gives the exact expression of $V_{2-1}$  as a function of the external electric field
and of the screening charge.

We now rewrite $D_a$ and $D_s$ as follows:
\begin{eqnarray}
\label{da}
D_a\ &=&\ d_a(n)\ E_{\rm{av}}, \\
D_s\ &=&\ \frac{\displaystyle e}{\displaystyle 2\epsilon_0}\ d_s(n)\ \Delta \rho,
\label{ds}
\end{eqnarray}
where $D_a$ has a linear dependence on the average electric field
through a proportionality constant $d_a$ which depends on the doping $n$.
$D_s$ is instead the contribution
to the interlayer polarization coming from the width of the transferred charge.
Therefore we write it in a form consistent
with the other interlayer term in Eq.(\ref{nuova-formuletta}), $i.e.$, $d e^2/(2\epsilon_0)  \Delta \rho$,
with a proportionality constant $d_s$ which depends on the doping $n$.

\begin{figure}[t]
  \centering
  \includegraphics[width=1.1\columnwidth]{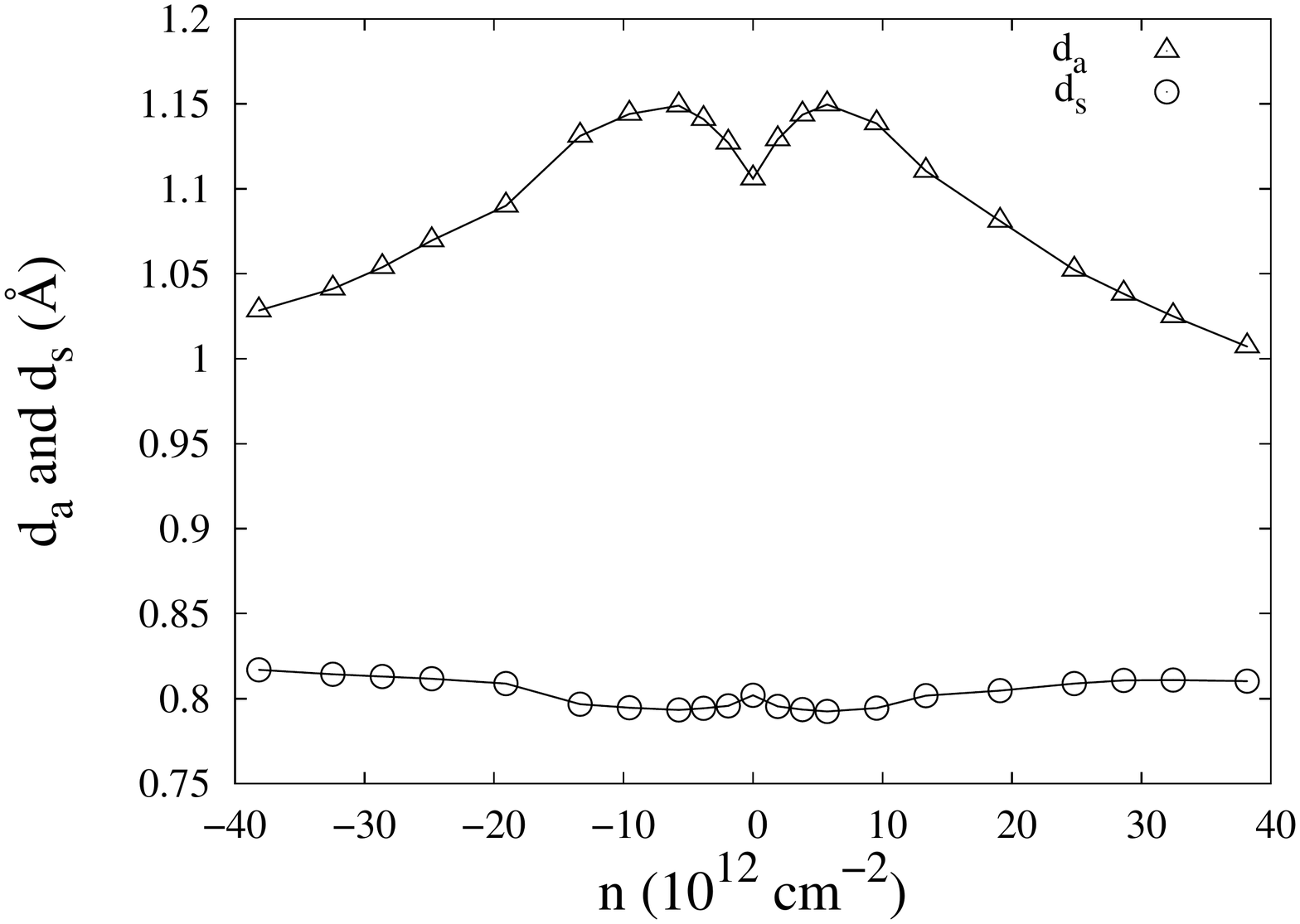}
  \caption{ $d_a$ and $d_s$ as a function of doping $n$,
            as defined in Eqs.(\ref{da}) and (\ref{ds}).
 }
  \label{ds-da}
\end{figure}

In Fig.\ref{ds-da} we show $d_a$ and $d_s$ as a function of doping $n$.
Since $d_s$ is almost independent of the doping and $d_a$ has a variation
in the order of 5\%, we replace them with their average values calculated on the doping range considered,
$\bar{d}_a$=1.09 \cf{\AA}, and $\bar{d}_s$=0.80 \cf{\AA}.
Using only this approximation and Eq.(\ref{beta}), we have
\begin{equation}
U =\beta \left[-e(d-\bar{d}_a) E_{\rm{av}} + \frac{\displaystyle e^2}{\displaystyle 2\epsilon_0} (d-\bar{d}_s)\Delta \rho\right].
\label{new-pot}
\end{equation}

We notice that the simplified electrostatic model
described in Sec.\ref{sec:com}, $i.e.$, Eq.(\ref{eq-for-TB}),
used in TB calculations,
is equivalent to consider, in Eq.(\ref{new-pot}),
$\beta = 1$, $\bar{d}_a = 0$ ($i.e.$ $\rho^{(1)}_a (z) = 0$),
and $\bar{d}_s = 0$ [$i.e.$, $\rho^{(1)}_s (z)= \delta (|z|-d/2)\ \Delta \rho /2\ \rm{sign}(z)$].

Considering Eq.(\ref{new-pot}), and the definition of $\alpha$ [as in Eq.(\ref{def-alphagap})]
 and $\eta$ [as in Eq.(\ref{def-eta})] we obtain another relation between $\alpha$ and $\eta$
as follows:
\begin{equation}
\alpha (n) = \frac{\alpha^{\cf{bare}}\beta\ (d-\bar{d}_a)/d }{1\ -\ \eta (n)\ \alpha^{\cf{bare}}\beta\ (d-\bar{d}_s)/d}.
\label{new-model}
\end{equation}

Equation (\ref{new-model}) gives the approximate relation between $U$, the average electric field,
the screening charge obtained considering the intralayer polarization,
and considering the width of the transferred charge between layers.
This equation substitutes Eq.(\ref{formuletta-wrong}) which comes from the simplified
electrostatic model described in Sec.\ref{sec:com}.

\begin{figure}[t]
  \centering
  \includegraphics[width=1.1\columnwidth]{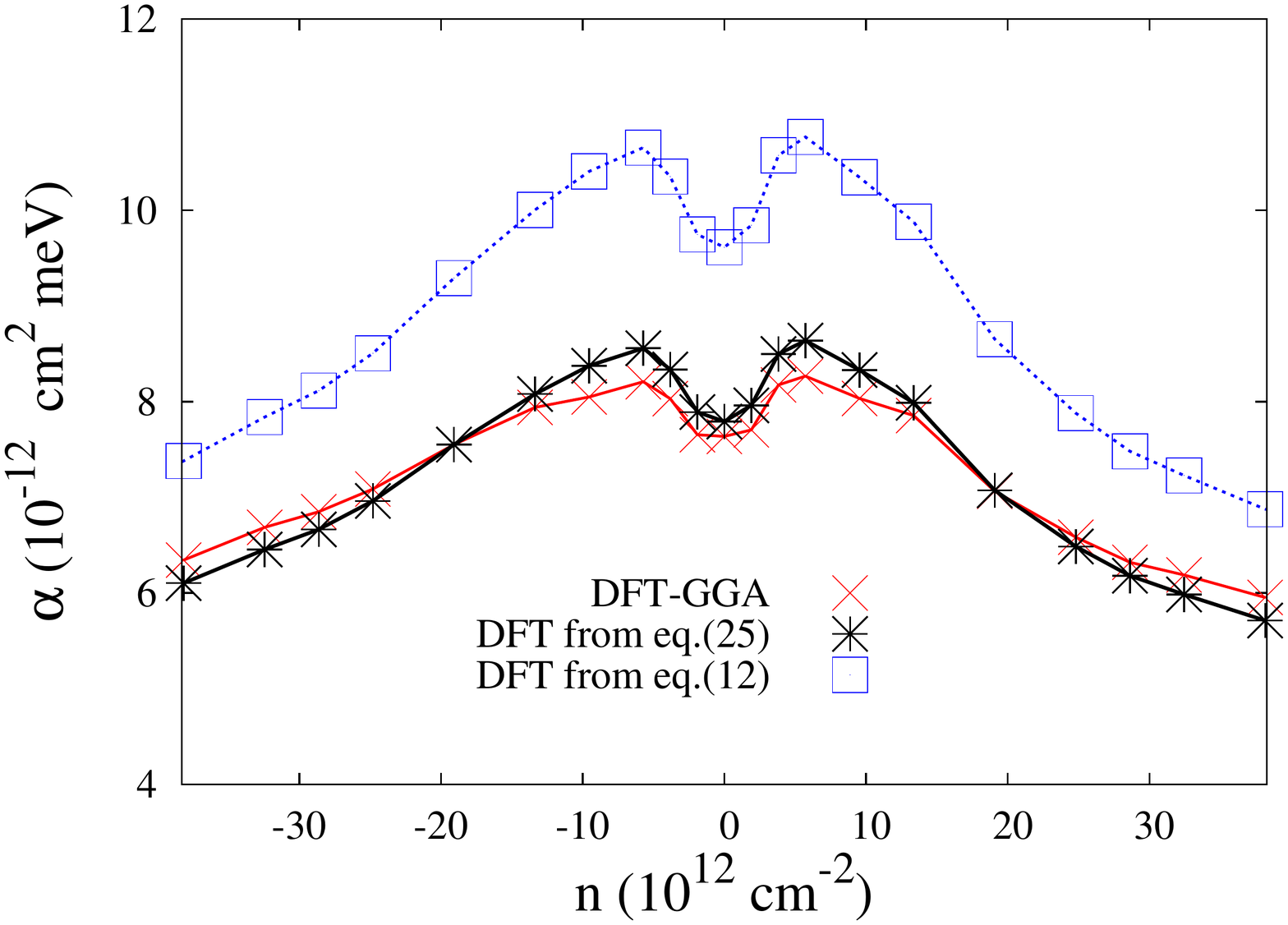}
 \caption{ (Color online) $\alpha(n)$ calculated with DFT-GGA ($\times$ crosses)
          from Eq.(\ref{formuletta-wrong}) using the DFT-GGA calculated $\eta(n)$ (squares)
          and from Eq.(\ref{new-model}) using the DFT-GGA calculated $\eta(n)$ (stars).
 }
  \label{our-model}
\end{figure}

In Fig.\ref{our-model} we show
the DFT-calculated $\alpha(n)$,
$\alpha(n)$ obtained from Eq.(\ref{formuletta-wrong}) using the
DFT-calculated $\eta(n)$,
and from the electrostatic model of Eq.(\ref{new-model}) using the
DFT-calculated $\eta(n)$.
One can see that the simplified electrostatic
model is not able to describe the DFT results.
Instead, $\alpha(n)$ obtained from the model of
Eq.(\ref{new-model}) is able to correctly reproduce
the DFT calculations.


\subsection{\label{Temperature} Effect of the electronic temperature on $\alpha$ as a function of doping $n$}

In Sec. \ref{sec:alpharho-U} we have shown that in the bilayer the electronic screening to
the external electric field is crucial for a correct evaluation of the band-gap.
At low doping level the screening is expected to depend on the broadening parameter,
and in this section we investigate the effect of the electronic temperature on the screening and on $\alpha$.

\begin{figure}[t]
  \centering
  \includegraphics[width=1.1\columnwidth]{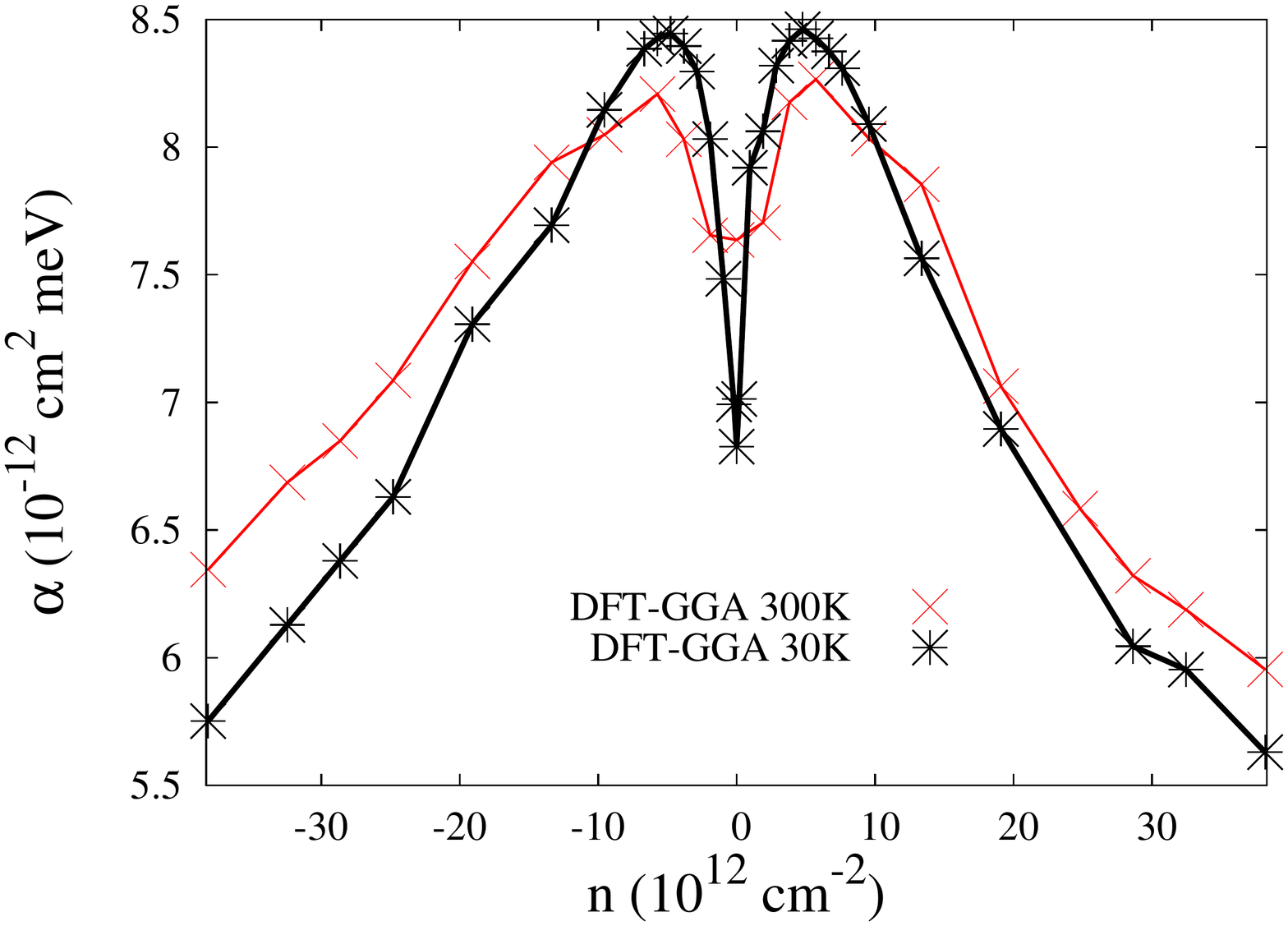}
  \caption{ (Color online) The DFT-GGA calculated $\alpha$  as a function of the doping
            $n$ for an electronic temperature of 300 (crosses)
           and 30 K (stars).
 }
  \label{temp-dep}
\end{figure}

In Fig.\ref{temp-dep} we show the DFT-calculated $\alpha$ as a function of $n$
for an electronic temperature of 300 and 30 K.
The variation in screening with the broadening parameter depends on the doping $n$.
Since the doping levels which are interesting for applications of the bilayer as active device in
nanoelectronics are small values around the zero doping, we focus on this doping range.

\begin{figure}[t]
  \centering
  \includegraphics[width=1.1\columnwidth]{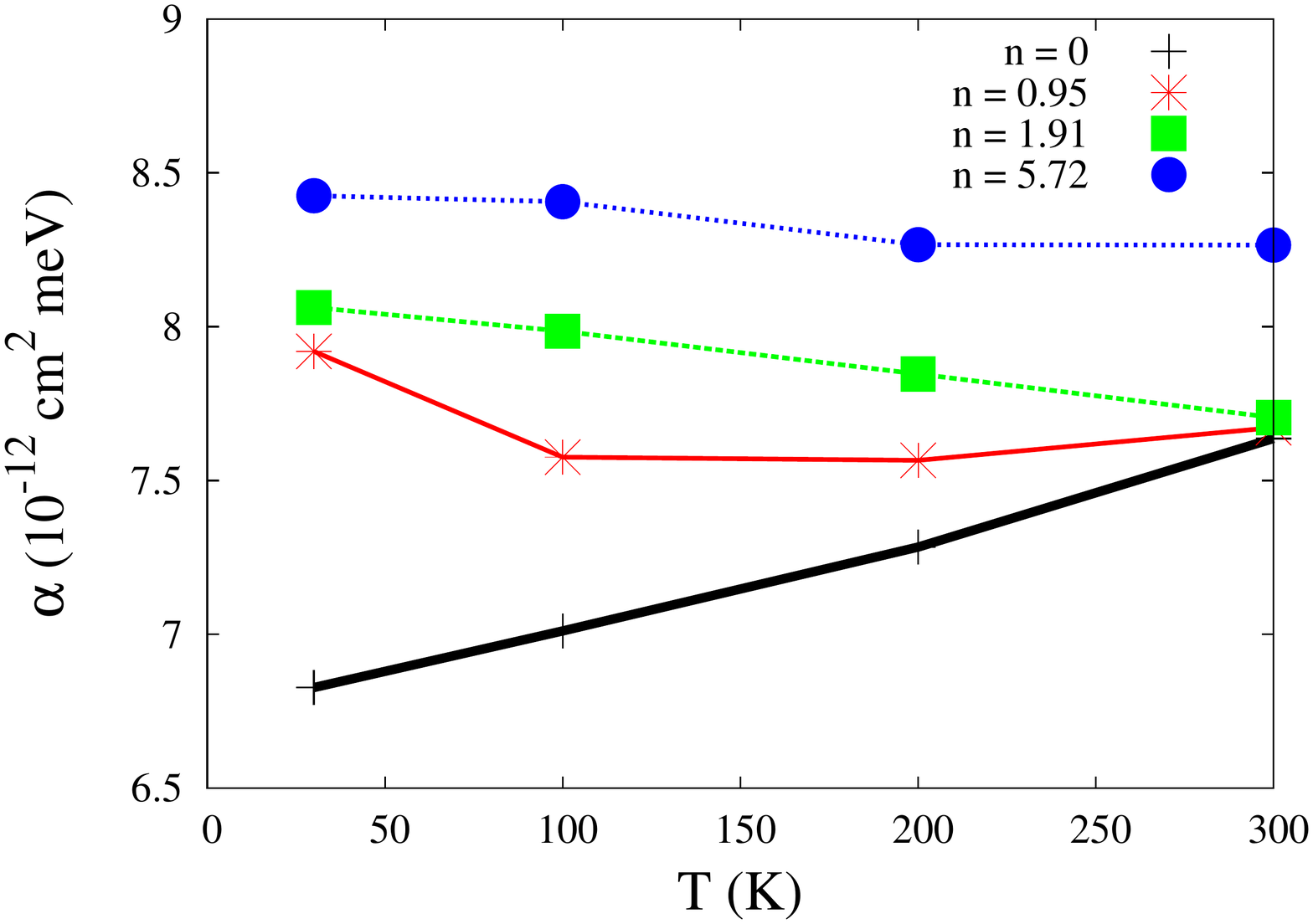}
  \caption{ (Color online) The DFT-GGA calculated $\alpha$ as a function of the electronic temperature $T$,
           for small values of electron doping. The values of doping $n$ are in units of \cf{10^{12} cm^{-2}}.
 }
  \label{varie-T}
\end{figure}

In Fig.\ref{varie-T} we show the DFT-calculated $\alpha$ as a function of the electronic temperature $T$
for electron doping values between 0 and 5.72 \cf{\times\ 10^{12}cm^{-2}}.
In this range of doping, we can see that the difference between $\alpha$
at 300 and 30 K is largest for zero doping.
In particular, for zero doping the band-gap at 30 K results to be about 10\% smaller than
at 300 K.


\subsection{\label{formuletta} Nonmonotonic behavior of $\alpha$
                as a function of doping $n$}

As shown in Figs.\ref{fig:alpha-gap} and \ref{fig:etaabinit}, both DFT-calculated $\alpha$
and $\eta$ have a nonmonotonic behavior as a function of the doping $n$.
$\alpha$ and $\eta$ represent the linear response of $U$ to
the external average electric field $E_{\rm{av}}$ and the linear response of $\Delta \rho$
to the band-gap $U$, respectively. Up to now, we calculated $\alpha$
and $\eta$ for finite values of $E_{\rm{av}}$ and $U$.
In this section we show that
perturbation theory (PT) explains the origin of this nonmonotonic behavior as  a function
of doping $n$.

For the numerical evaluation of the expressions obtained from PT, we use the band structure
calculated with the TB model.
Indeed, even if TB results for $\alpha$ and $\eta$ differ from the DFT ones,
TB is able to catch the nonmonotonic trend of these quantities
as a function of the doping $n$.
Moreover, we limit our PT calculations to $\eta (n)$. Indeed, since the relation between $\eta$ and $\alpha$
is monotonic [see Eq.(\ref{new-model})], the nonmonotonic behavior of $\eta$
is able to explain also the nonmonotonic behavior of
$\alpha$ as a function of $n$.

In order to calculate $\eta (n)$ with PT,
we consider $\hat{H}^{(0)}_{\bf{k}}$, which is the unperturbed TB Hamiltonian.
$\hat{H}^{(0)}_{\bf{k}}$
is a $4\times 4$ matrix which depends on the wave vector $\bf{k}$ and is
written on the basis of $2p_z$ orbitals centered on the four atoms of the
unit-cell, ordered as $A$, $B$, $A'$, $B'$
($A$ and $B$ are the two carbon atoms on layer 1,  $A'$ and $B'$ are the two
carbon atoms on layer 2,
and in the Bernal stacking configuration $A$ and $A'$ are vertically superposed).
In presence of a band splitting $U$ (see Fig.\ref{bande}),
the Hamiltonian $\hat{H}_{\bf{k}}$ can be written as
\begin{equation}
\hat{H}_{\bf{k}}=\hat{H}^{(0)}_{\bf{k}}+\frac{U}{2} \widehat{\Delta \rho},
\end{equation}
where
\begin{equation}
\widehat{\Delta \rho}=
\left(\begin{array}{cccc}
1 & 0 & 0 & 0 \\
0 & 1 &  0 & 0 \\
0 & 0 & -1 & 0 \\
0 & 0 & 0  & -1
\end{array} \right).
\label{pippoequation}
\end{equation}

Using first-order PT,
 we obtain the following expression for $\eta = \left(\frac{\displaystyle d\Delta \rho}{\displaystyle dU}\right)$:

\begin{eqnarray}
\eta = \frac{1}{N_k} \sum_{\mathbf{k},i,j\neq i}
&\frac{\displaystyle f(\epsilon_{i\mathbf{k}}^{(0)}-\epsilon_{F}^{(0)})-f(\epsilon_{j\mathbf{k}}^{(0)}-\epsilon_{F}^{(0)})}{\displaystyle \epsilon_{i\mathbf{k}}^{(0)}-\epsilon_{j\mathbf{k}}^{(0)}}&\ \times \nonumber \\
&\times\ \vert <\psi^{(0)}_{j\mathbf{k}}\vert \widehat{\Delta \rho} \vert \psi^{(0)}_{i\mathbf{k}}>\vert^2 ,&
\label{PT+TB}
\end{eqnarray}
where $|\psi^{(0)}_{i\mathbf{k}}>$ ($i$=1,2,3,4)
are the unperturbed eigenstates of $\hat{H}^{(0)}_{\mathbf{k}}$ with eigenvalues $\epsilon_{i\mathbf{k}}^{(0)}$,
$\epsilon_{F}^{(0)}$ is the Fermi level, $f(\epsilon_{i\mathbf{k}}^{(0)}-\epsilon_{F}^{(0)})$ is the occupation
of state $i$, and $N_k$ is the number of
$\mathbf{k}$ points used in the BZ integration.

\begin{figure}[t]
  \centering
  \includegraphics[width=0.9\columnwidth]{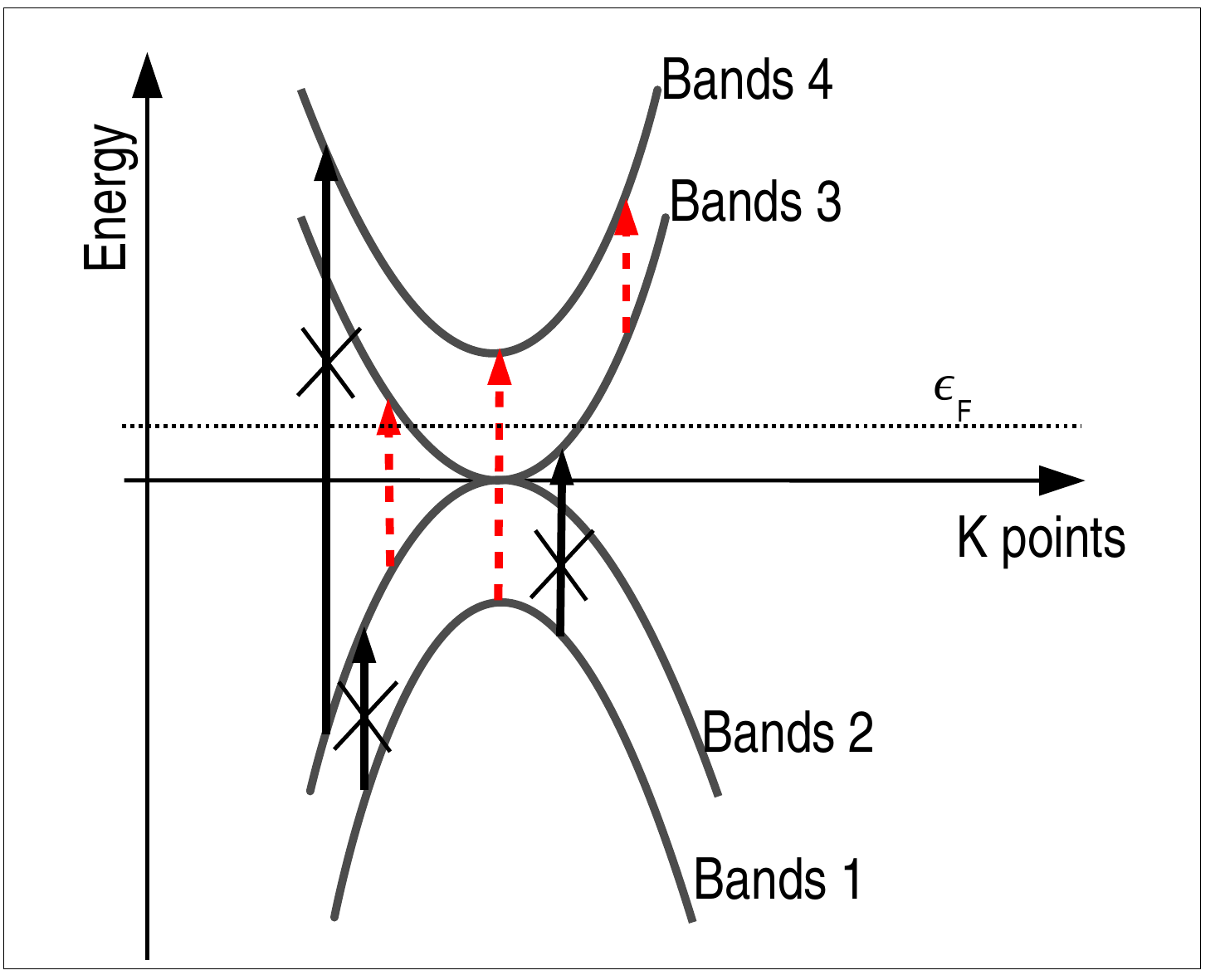}
  \caption{(Color online) Schematic representation of allowed (dashed arrows) and not allowed (continuous arrows)
       contributions from band-to-band transitions in Eq.(\ref{PT+TB}), for electron doping.
 }
  \label{bands-scheme}
\end{figure}

In particular, in Eq.(\ref{PT+TB}) there are six contributions obtained by mixing the
four unperturbed states $|\psi_{i\mathbf{k}}^{(0)}>$, with $i$=1,2,3, and 4; and we label
as $\eta_{i,j}$ the contribution to $\eta$ given by
states $i$ and $j$.
Within the present TB model $\eta_{1,3}$ and $\eta_{2,4}$ are exactly zero. \cite{note_pt}
Contribution $\eta_{1,2}$ vanishes for $\epsilon_F > 0$ because the two states are
both occupied.
Therefore, for $\epsilon_F > 0$, the important contributions to $\eta$
 derive from $\eta_{1,4}$, $\eta_{2,3}$, and $\eta_{3,4}$,
as schematically shown in Fig.\ref{bands-scheme}.

\begin{figure}[t]
  \centering
  \includegraphics[width=1.1\columnwidth]{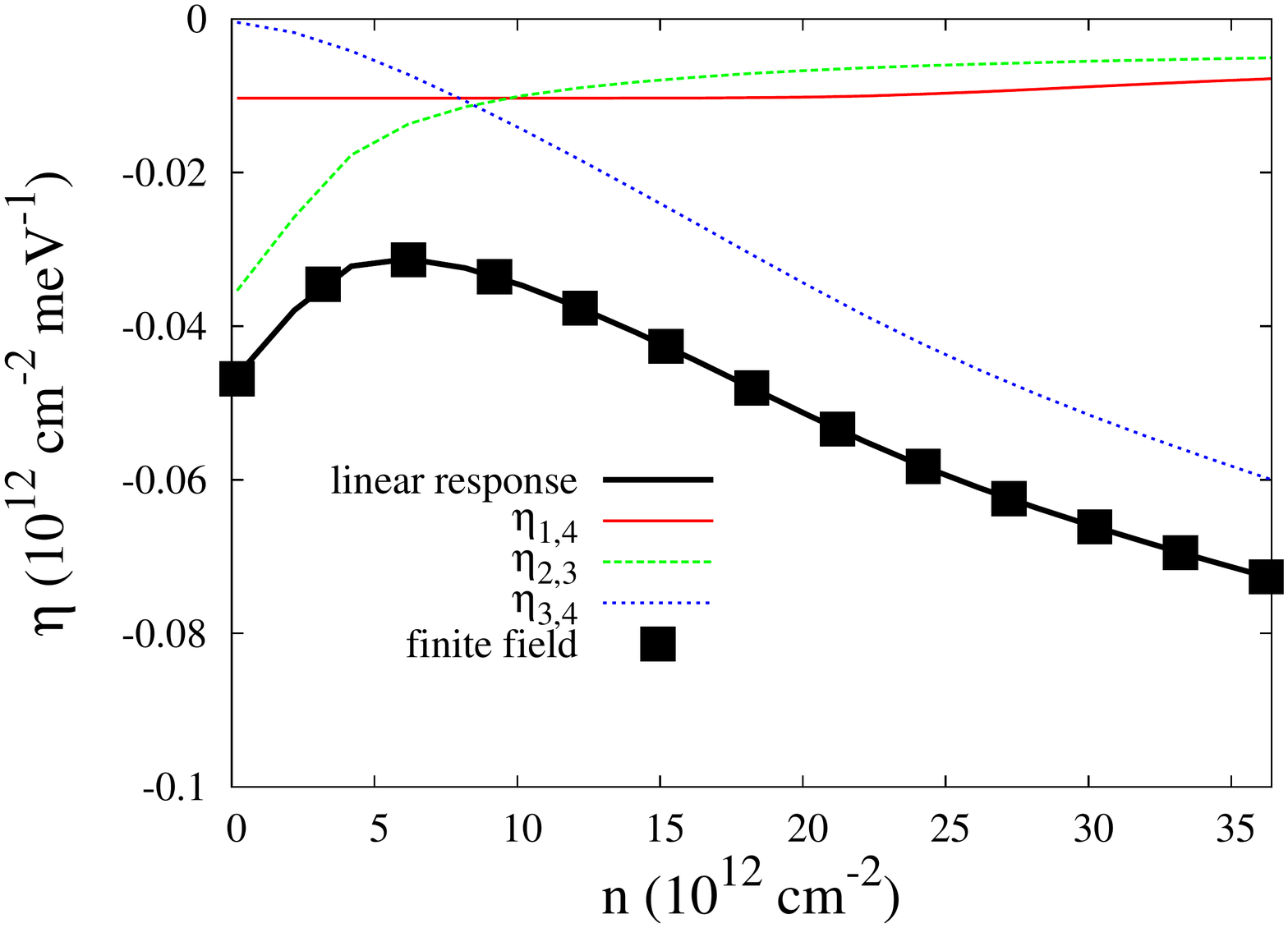}
  \caption{ (Color online) $\eta$ as a function of electron doping, from Eq.(\ref{PT+TB}) (linear response),
            and different contributions from $\eta_{1,4}$, $\eta_{2,3}$, and $\eta_{3,4}$.
            Squares are the nonperturbative TB results.
 }
  \label{eta}
\end{figure}

In Fig.\ref{eta} we show $\eta$ as a function of the electron doping,
obtained from Eq.(\ref{PT+TB}), for
an electronic temperature of 300 K. Different contributions from $\eta_{1,4}$, $\eta_{2,3}$, and
$\eta_{3,4}$ are also plotted.
For comparison, we report $\eta (n)$ calculated with the nonperturbative TB model.
Contribution $\eta_{1,4}$ as a function of doping is constant when the Fermi level
is lower than the bottom of band 4, and its absolute value starts to decrease
when band 4 becomes occupied due to lower availability of empty states.
Contribution $\eta_{2,3}$ is minimum for zero doping, and its absolute values decreases,
as a function of the electron doping, for the same reason. Contribution from $\eta_{1,4}$ is lower than contribution
from $\eta_{2,3}$ due to the energy difference in the denominator of Eq.(\ref{PT+TB}), which is higher for
$\eta_{1,4}$.
Finally, contribution  $\eta_{3,4}$ vanishes for zero doping, and its absolute value increases with
increasing electron doping, due to larger number of possible transitions between occupied and unoccupied
states.

The results  in Fig.\ref{eta} show that the nonmonotonic behavior of $\eta (n)$ is
determined by the sum of the three contributions $\eta_{1,4}$, $\eta_{2,3}$, and $\eta_{3,4}$,
which gives a maximum at the electron doping values $n \approx$ \cf{6\times\ 10^{12}\ cm^{-2}}.


\subsection{\label{GW}$GW$ correction}

Recently it has been shown by ARPES measurements that the electronic band structure
of graphene and graphite is not well reproduced  by
 LDA and GGA. \cite{Att_PRL100}
In particular, LDA and GGA underestimate the slope of the bands
since these approximations do not include long-range
electron-correlation effects. Such effects can fully be taken into account within the $GW$ approach
[where the self-energy
is computed from the product of the electron Green's function ($G$) and the
screened Coulomb interaction ($W$)],
which is considered to be the most accurate first principles approach for the
electronic band structure. \cite{review-GW}
The $GW$ band structures for graphite and graphene are indeed in very good agreement with the
ARPES measurements. \cite{Att_cm}
In absence of an external electric field,
the DFT-calculated bands of bilayer graphene need to be scaled in order to reproduce
the $GW$-correct bands as
\begin{equation}
\epsilon_{i\mathbf{k}}^{GW} = \lambda \ \epsilon_{i\mathbf{k}}^{\rm{DFT}},
\label{GW-eigenvalues}
\end{equation}
where $\lambda = 1.18$ is the scaling factor, as obtained from Ref.\cite{Att_cm}.
Such scaling factor can change the screening properties of the bilayer, and in this section
we include it in our theoretical results.

If we focus on the quantity $\eta$, we can use the perturbative expression in Eq.(\ref{PT+TB}).
In such expression, we correct the DFT eigenvalues $\epsilon^{(0)}_{i\mathbf{k}}$ using Eq.(\ref{GW-eigenvalues}),
and we can neglect the $GW$ correction to the matrix elements
$<\psi^{(0)}_{j\mathbf{k}}\vert\ \widehat{\Delta \rho}\ \vert \psi^{(0)}_{i\mathbf{k}}>$
since it is commonly found that the DFT error in the wave functions is usually negligible
with respect to the error on the eigenvalues.
Within this approximation, it is easy to show that
\begin{equation}
\eta^{GW}(\epsilon_{F}, T)\ =\ \frac{1}{\lambda}\ \eta^{\rm{DFT}} (\frac{\epsilon_{F}}{\lambda}, \frac{T}{\lambda}),
\end{equation}
where $T$ is the temperature.
The computed $\eta^{GW}$ is shown in Fig.\ref{fig:etaabinit}.

$\alpha^{GW}$ can be computed from $\eta^{GW}$ using our model in Eq.(\ref{new-model}).
In order to minimize the error from our model, we write
\begin{equation}
\alpha^{GW} (n)\ =\ \frac{\alpha^{\cf{bare}} \beta (d-\bar{d}_a)/d}{1 -\ \eta^{GW}(n)\ \alpha^{\cf{bare}} \beta (d-\bar{d}_s)/d }\ + \ \Delta ,
\end{equation}
where
\begin{equation}
 \Delta = \alpha^{\rm{DFT}} \ - \frac{\alpha^{\cf{bare}} \beta (d-\bar{d}_a)/d}{1 -  \eta^{\rm{DFT}}\ \alpha^{\cf{bare}} \beta (d-\bar{d}_s)/d },
\end{equation}
gives an estimate of the error in Eq.(\ref{new-model}).
The computed $\alpha^{GW} (n)$ is shown in Fig.\ref{fig:alpha-gap},
and for low doping levels it is around 10\% higher than the DFT value.


\subsection{\label{forexpsec} Full band structure of gated bilayer graphene}

\begin{figure}[t]
  \centering
  \includegraphics[width=1.1\columnwidth]{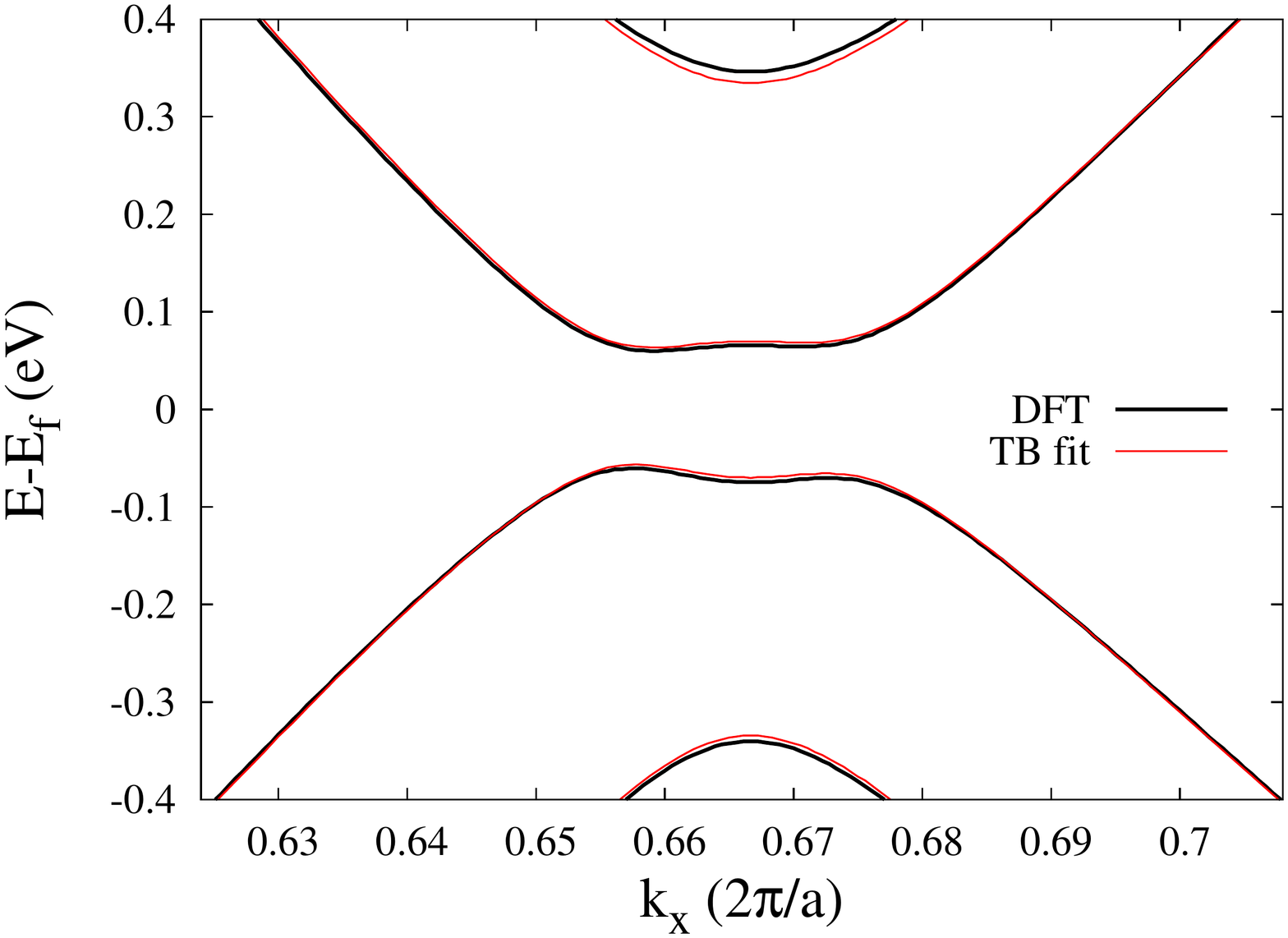}
  \caption{ (Color online) DFT-GGA calculated bands around the $K$ point in the BZ (DFT)
           for $n=0$ and with $U=0.14$ eV
           and from the TB model (TB fit), with five nearest-neighbors in-plane hopping parameters
           and three out-of-plane hopping parameters. }
  \label{bande-fit}
\end{figure}

In this section we give a practical instruction to obtain the full
band structure of bilayer graphene for a doping $n$ and for an
average electric field $E_{\rm{av}}$.
In order to do that, we fit our DFT bands along all the
$\Gamma KM$ line in the BZ in absence of the external electric
field, using a TB model with five nearest neighbors in-plane hopping
parameters ($\gamma_{\parallel}^1, \gamma_{\parallel}^2,
\gamma_{\parallel}^3, \gamma_{\parallel}^4, \gamma_{\parallel}^5$)
and three out-of-plane hopping parameters ($\gamma_{\perp}^{AA'},
\gamma_{\perp}^{AB'}, \gamma_{\perp}^{BB'}$). In the Bernal
stacking configuration of bilayer graphene, $A$ and $A'$ represent the
vertically superposed atoms. These hopping parameters do not
change when an external average electric field is applied. This is
shown in Fig.\ref{bande-fit}, where we compare the direct DFT
results with the TB calculations, with the fixed hopping parameters
and the $U$ value from the DFT calculations.

\begin{table}[t]
\footnotesize
\caption{TB-$GW$ parameters obtained by fitting the bilayer DFT bands with a TB model, along all the $\Gamma$KM line,
and by rescaling the parameters with $\lambda = 1.18$.
$\gamma_{\parallel}^i$ is the $i$-nearest-neighbors
hopping parameters. All values are in eV.}
\begin{tabular}{|c|c|c|c|c|c|c|c|c|}
\hline
  & $\gamma_{\parallel}^1$ &  $\gamma_{\parallel}^2$ & $\gamma_{\parallel}^3$ & $\gamma_{\parallel}^4$ & $\gamma_{\parallel}
^5$ & $\gamma_{\perp}^{AA'}$ & $\gamma_{\perp}^{AB'}$ & $\gamma_{\perp}^{BB'}$  \\
\hline
\hline
TB-$GW$ & -3.4013 & 0.3292 & -0.2411 & 0.1226 & 0.0898 & 0.3963 & 0.1671 & 0.3301 \\
\hline
\end{tabular}
\label{par-5nn}
\end{table}

\begin{table}[t]
\caption{ Values of fitting parameters of Eq.(\ref{fit-lorent}).
          All values are in $\rm{10^{-12}\ cm^2 meV }$.
}
\begin{tabular}{|c|c|}
\hline
$A_1$ & 0.896 \\
$B_1$ & -26.888 \\
$\gamma_1$ & 21.756 \\
\hline
$A_2$ & 3.905 \\
$B_2$ & 1.623\\
$\gamma_2$ & 21.946 \\
\hline
$A_3$ & -1.654 \\
$B_3$ & -0.092\\
$\gamma_3$ & 5.534\\
\hline
C & 5.848 \\
\hline
\end{tabular}
\label{parameters}
\end{table}

Since we consider the $GW$ one as the most precise result,
in Table \ref{par-5nn} we report the TB-$GW$ hopping parameters obtained by fitting the
DFT bands without electric field and by rescaling them with the $GW$ scaling factor $\lambda=1.18$.
Moreover, in order to avoid the numerical evaluation of $U$ for a given $n$ and $E_{\rm{av}}$,
we give a fit of our calculated $\alpha^{GW}(n)$:
\begin{equation}
\alpha^{GW}(n) = \sum^{3}_{i=1} \frac{\displaystyle A_i}{\left[1+\frac{\displaystyle (n-B_i)^2}{\displaystyle \gamma_{i}^2}\right]} + C,
\label{fit-lorent}
\end{equation}
where the values of the fitting parameters are listed in Table \ref{parameters}.
In Fig.\ref{fig:alpha-gap} we show the results of the fit with the black continuous line.
From expression (\ref{fit-lorent}), we can obtain the value of the gap $U$
as a function of the doping $n$ and of the external average electric field $E_{\rm{av}}$,
\begin{equation}
U(n,E_{\rm{av}})\ =\ \alpha^{GW}(n) (n_1 - n_{2}),
\label{forexp}
\end{equation}
where $(n_1 - n_{2}) = E_{\rm{av}}/ (|e|/(2\epsilon_0))$.
$n$, $n_1$, and $n_2$ are in units of \cf{10^{12} cm^{-2}}.


\subsection{\label{sec:exp}Comparison with experimental results}

\begin{figure}[t]
  \centering
  \includegraphics[width=1.1\columnwidth]{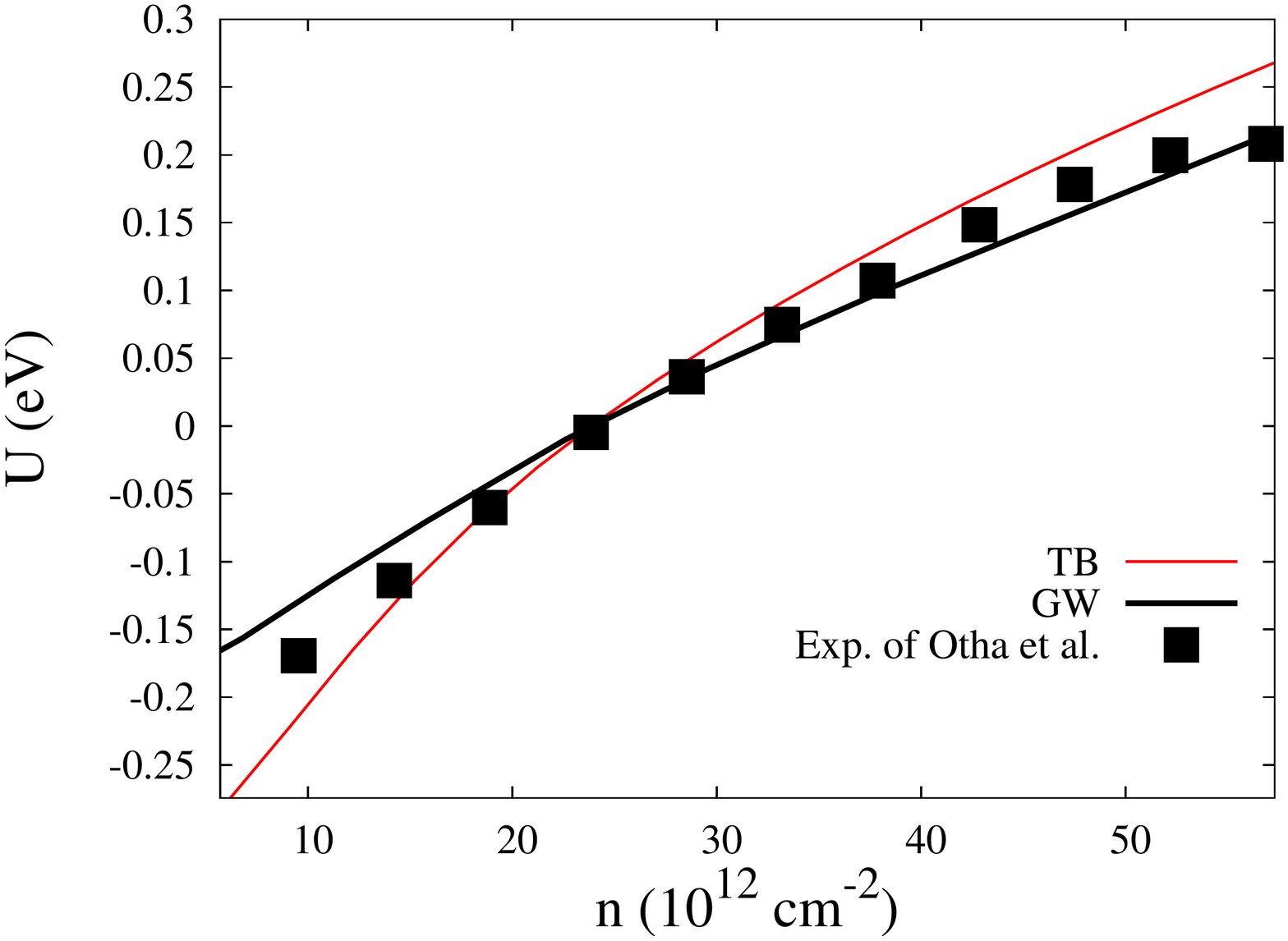}
  \caption{ (Color online) Comparison between experimental results for $U$ from Ref.\cite{Ohta_science} (squares),
           TB, and $GW$  results.
 }
  \label{science}
\end{figure}

In this section we compare our DFT and TB results
with the direct measurement (ARPES) of the band-gap on epitaxially growth bilayer graphene in
 Ref.\cite{Ohta_science}.
In this work Ohta $et\ al.$\cite{Ohta_science} performed an experiment where bilayer
graphene is synthesized on silicon carbide (SiC) substrate. The
SiC acts as a fixed bottom gate, and a charge $n_2$ flows from the
substrate to the bilayer. Further electron doping is induced
with the deposition of potassium atoms on the other side of the
bilayer, and this chemical doping acts as a top gate. Varying the
concentration of potassium, the asymmetry between the two layers
of graphene is modified, and a band-gap is opened accordingly.
Using angle-resolved photoemission spectroscopy Ohta $et\ al.$\cite{Ohta_science}
directly measured the band structure, and fitting it with a TB
model, they obtained a curve of the gap as a function of the doping
charge in the bilayer.

To compare with their experimental results,
we calculate the gap using $\alpha^{GW} (n)$ from Eqs.(\ref{fit-lorent}) and (\ref{forexp})
and keep $n_2$ (bottom gate) fixed at 11.9 \cf{\times\ 10^{12} cm^{-2}}.
This value of $n_2$ derives from the fact that in Ref.\cite{Ohta_science},
for a total doping of $n = 23.8$ \cf{\times\ 10^{12} cm^{-2}}, no gap is observed,
meaning that $n_1 = n_2 = n/2$. Since in the experiment the bottom gate is not varied,
we also keep it fixed to this value, and we only vary $n_1 = n - n_2$.

In Fig.\ref{science} we compare our results, obtained with $\alpha^{GW}$
and with $\alpha^{\rm{TB}}$, with the experimental data from Ref.\cite{Ohta_science}
We first notice that the nonlinearity is not due to the saturation of the gap with $E_{\rm{av}}$;
it is instead due to the dependence of $\alpha$ on the doping $n$
(at high doping $\alpha$ decreases with n).
Moreover, both $GW$ and TB results are in good agreement with experiments.
This is due to the fact that the experiment is carried out at high doping levels,
where the difference between the $GW$ and TB $\alpha$'s is less important with respect to
low doping levels (see Fig.\ref{fig:alpha-gap}).

In the case of exfoliated bilayer graphene, direct experimental measurements of the band structure
and of the gap with ARPES are still unavailable.
Alternatively, indirect information on the band structure can be obtained by
infrared reflectivity studies.
Recently, Kuzmenko $et\ al.$\cite{Kuzmenko-infrared} reported
 an experimental work on infrared spectra of exfoliated and gated bilayer graphene as a function of
doping.
In this work the
authors found a strong gate-voltage dependence of their spectral features, which are related
to interband transitions.  A comparison of the experimental infrared spectra
with the one obtained from TB calculations suggests that
the TB prediction of gate-induced band-gap is overestimated.\cite{Kuzmenko-infrared}
However, a quantitative
analysis of the band-gap as a function of doping and external field is not given.

Finally, by measuring the cyclotron mass as a function of doping in bilayer graphene
one can check the presence of a finite band-gap.
These measurements do not provide a direct estimate of the band-gap; however, they give important
informations on the hole-electron asymmetry and on the deformation of the band
structure in the presence of an external electric field.
In particular, in Refs.\cite{Castroneto_PRL} and \cite{Castroneto_arXiv} the authors measured the cyclotron mass
on exfoliated bilayer graphene.
The bottom gate is realized with an oxidized silicon substrate,
which allows a variation in bottom gate electron density $n_{2}$ during the experiment.
The top gate is provided by chemical doping,
by deposition of \cf{NH_3} molecules, which provides a top gate electron density
$n_{1}$, which is fixed during the experiment.

To compare the experimental results of Refs.\cite{Castroneto_PRL} and \cite{Castroneto_arXiv} with
our band structures, we calculate
the cyclotron mass $m_c$ as
\begin{equation}
m_c (n) = \frac{\hbar^2}{2 \pi}\ \left(\frac{dA(E)}{dE}\right)_{E=E_f(n)},
\label{cyclotronm}
\end{equation}
where $A$ is the $\mathbf{k}$-space area enclosed by the orbit with energy $E$ and $E_f$ is the Fermi level.
The derivative in Eq.(\ref{cyclotronm}) is obtained by finite differentiation with respect to $E$.
For the $GW$ calculations we use the band structure calculated as described in Sec.~\ref{forexpsec}.

\begin{figure}[t]
  \centering
  \includegraphics[width=1.1\columnwidth]{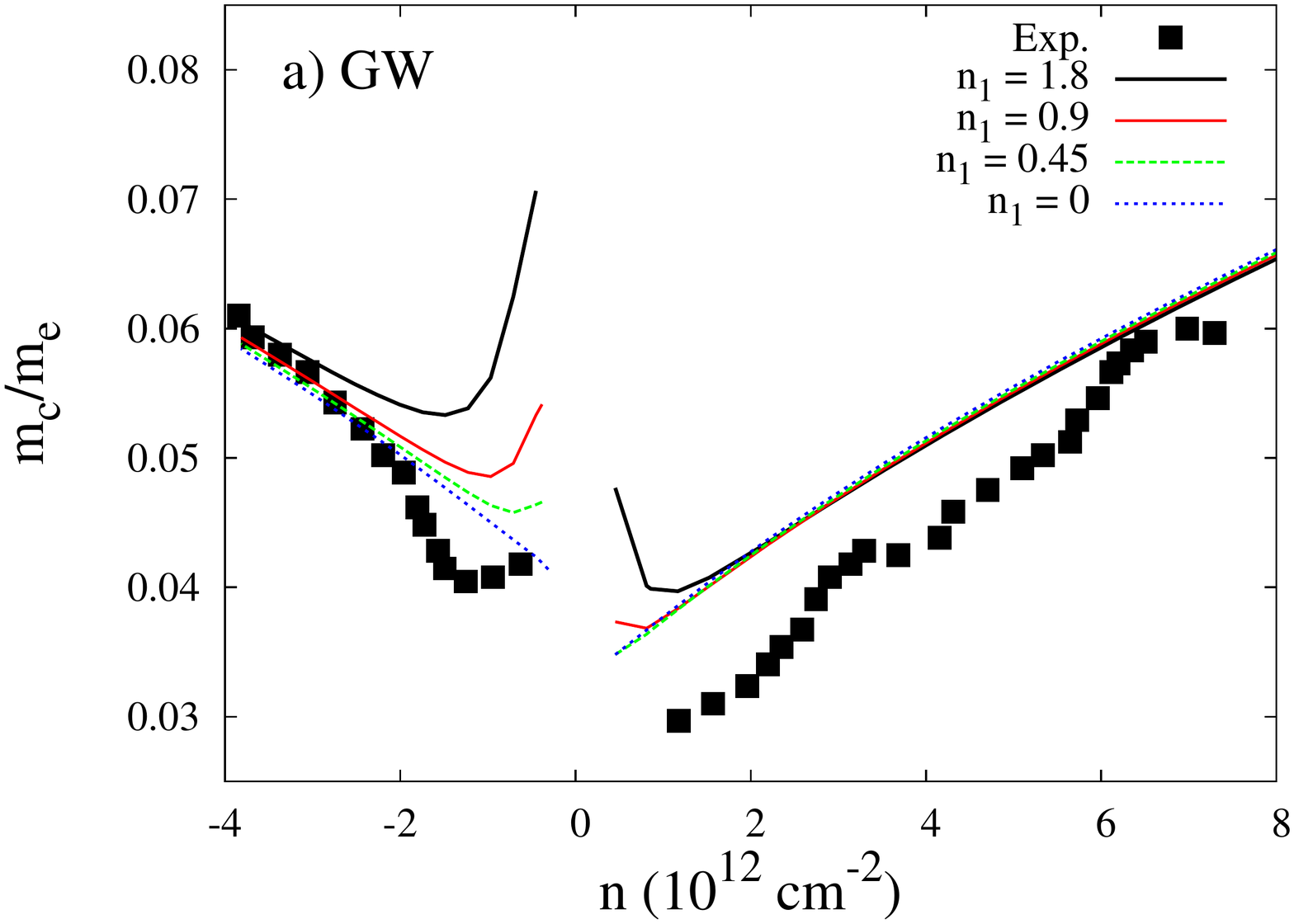}\\
  \includegraphics[width=1.1\columnwidth]{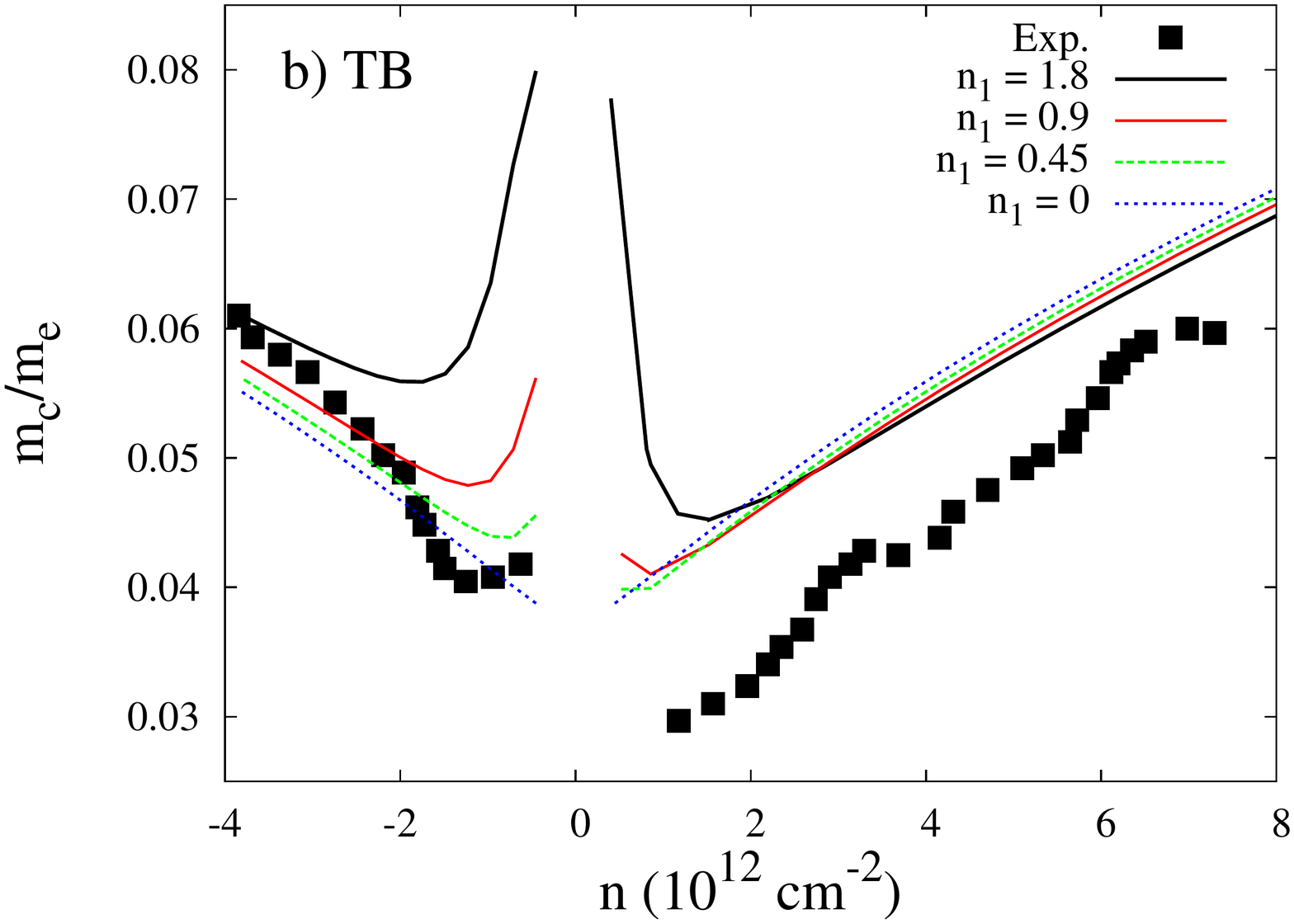}
  \caption{(Color online) Cyclotron mass, with respect to the free electron
           mass $m_{e}$, as a function of doping $n$:
           comparison of the experimental results
           from Ref.\cite{Castroneto_PRL}
           with our (a) $GW$ calculations and with our (b) TB calculations for different values of $n_{1}$
           (the values of $n_1$ are in units of \cf{10^{12} cm^{-2}}).
 }
  \label{mc}
\end{figure}

In Fig. \ref{mc} we compare the experimental data on the cyclotron
mass from Ref.\cite{Castroneto_PRL} with our [Fig.\ref{mc}(a)] $GW$ calculations
and with [Fig.\ref{mc}(b)] TB calculations\cite{parametro} for
different values of top gate electron density $n_{1}$. In
Ref.\cite{Castroneto_PRL} the authors estimated an initial doping
$n_0$ on bilayer graphene, at zero bottom gate, of about  1.8
\cf{\times 10^{12} cm^{-2}}. In principle, such initial doping
could come both from the deposited \cf{NH_3} molecules ($i.e.$,
from the top gate) and from a charge transfer from the \cf{SiO_2}
substrate ($i.e.$, from the bottom gate). Thus an exact estimation
of the top gate electron density $n_{1}$ is not possible, and we
calculate the cyclotron mass for values of $n_{1}$ between 1.8
\cf{\times 10^{12} cm^{-2}} and 0. Our results show that for both
$GW$ and TB calculations, the cyclotron mass behavior as a function
of doping depends on the value of $n_{1}$. In particular, for the
$GW$ calculations the best agreement with the experimental results
is obtained for $n_{1}$ = 0.45~\cf{\times 10^{12} cm^{-2}}.
Finally, we note that our $GW$ calculations give better results than
the TB ones. In particular, contrary to the TB results, they are
able to reproduce the hole-electron asymmetry.


\section{\label{concl}Conclusions}

We present a detailed ab initio DFT investigation of the band-gap
opening and screening effects of gated bilayer graphene. First, we
analyze the response of the band-gap to the external average
electric field at fixed doping. We show that this response is
linear for different electron and hole doping values and for
large electric field values. We then find that the linear response
of the gap to the electric field has a nonmonotonic behavior as a
function of doping and for low doping values it depends on the
temperature.

We also perform TB calculations for the band-gap opening. At low
doping values, which are the interesting ones for electronic
applications, we find that the DFT-calculated gap is roughly half
of the TB one. Since the band-gap strongly depends on the
screening effects, we perform a detailed analysis of the charge
distribution in the bilayer in presence of the external electric
field. We show that the electronic screening is characterized by
interlayer and intralayer polarizations. The latter one, not
included in TB calculations, gives an important contribution to
the band-gap opening.

On the basis of this analysis, we propose a model which
significantly improves the description of the electronic
properties of bilayer graphene in the presence of an external
electric field, and finally we provide a practical scheme to
obtain the full band structure of gated bilayer graphene for
arbitrary values of the doping and of the external electric field.


\section*{ACKNOWLEDGMENTS}
Calculations were performed at the IDRIS supercomputing center
(Project No. 081202 and 081387).


\appendix

\section{\label{dip-mon}Dipole and monopole potential}

\begin{figure}[t]
  \centering
  \includegraphics[width=1.1\columnwidth]{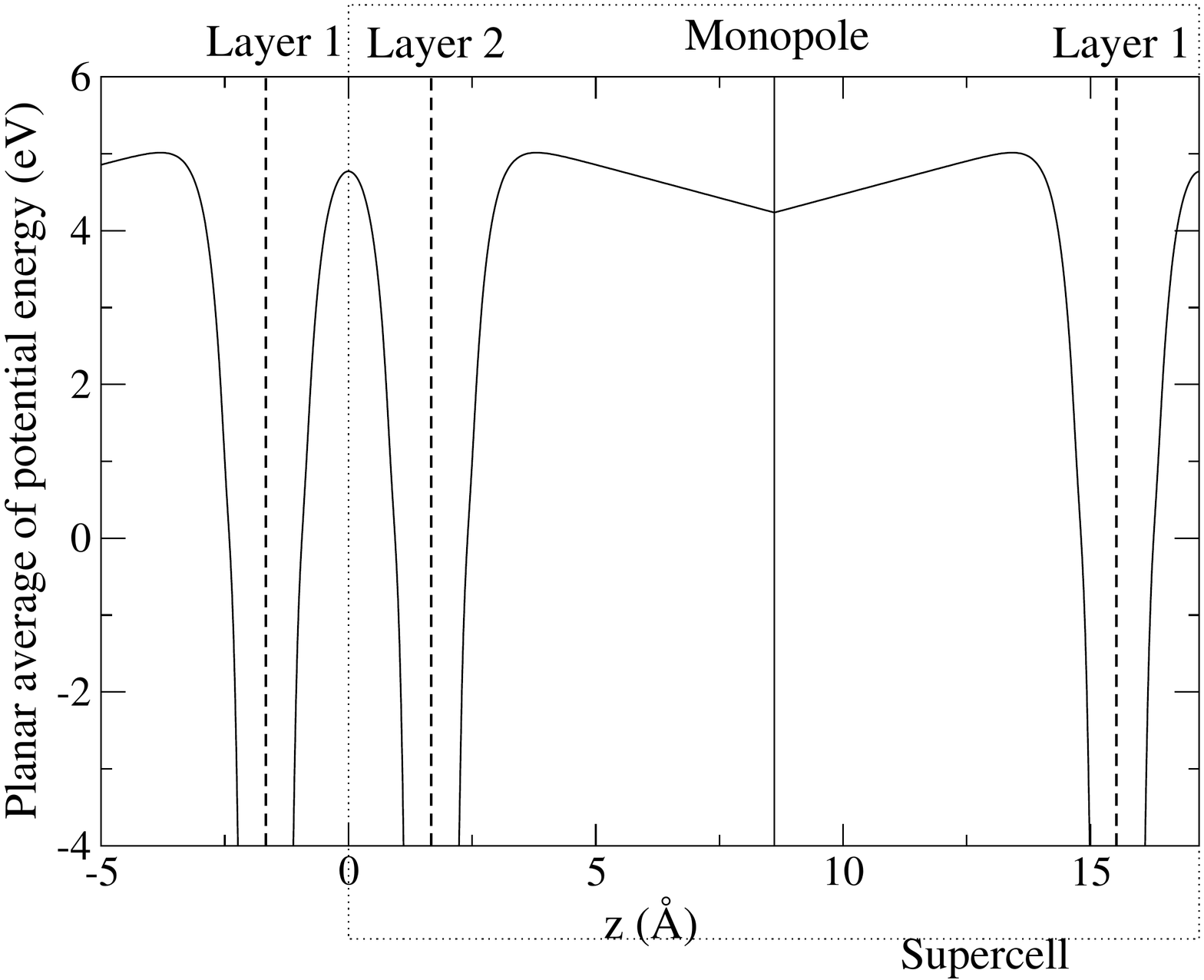}
  \caption{Planar average of the ionic, Hartree, and monopole potentials,
multiplied by the electron charge.
This figure corresponds to
a doping charge on the bilayer $n$ = \cf{19\times 10^{12}\  cm^{-2}}.
The positions of the first and second layers of the bilayer and of the monopole
in the supercell are indicated.}
  \label{fig:pot-mono}
\end{figure}

Standard plane-wave ab initio codes work with
periodically repeated super-cells.
When doping the sample with a total
charge $neA$ ($A$ is the area of the section of the periodic cell
parallel to the graphene plane)
a compensating uniform background charge (with opposite sign)
is added in order to have a neutral system and, thus, a
periodic electrostatic potential.

In this work, we use a different approach, and
we add a "monopole", that is a uniformly charged
plane equidistant from the two graphene layers, with total charge
$-neA$.
This is done by adding in real space a periodic potential energy
given by
\begin{equation}
V_{\rm{mon}} (\bar{z}) = -\frac{ne^{2}}{2\epsilon_0}\  (-\bar{z} + \frac{\bar{z}^2}{L}),
\end{equation}
where $\bar{z} = z - z_{\rm{mon}}$, $z_{\rm{mon}}$ is the $z$ coordinate of the monopole plane,
and $\bar{z} \in [0;L]$, where $L$ is the length of the supercell along $z$.
While the linear term of $V_{\rm{mon}}$ is the potential associated with the
monopole plane, the quadratic term cancels the potential associated
with uniform background.
$V_{\rm{mon}}$ can be added to the electrostatic potential acting on the
Kohn-Sham electronic states with a straightforward implementation.
The resulting system is, as a whole, neutral and periodic.

In Fig.~\ref{fig:pot-mono} we show the planar average of the ionic, Hartree, and monopole
potentials, multiplied by the electron charge.
The position of the first and second layers of the bilayer in the supercell is indicated,
together with the monopole position.
The distance between the monopole and the bilayer is 6.93 \cf{\AA}.
This figure corresponds to a doping charge on the bilayer $n$ = \cf{19\times 10^{12}\  cm^{-2}}
and to an experimental setup where the bottom and top gates are equal,
and no gap opening is expected.

\begin{figure}[t]
  \centering
  \includegraphics[width=1\columnwidth]{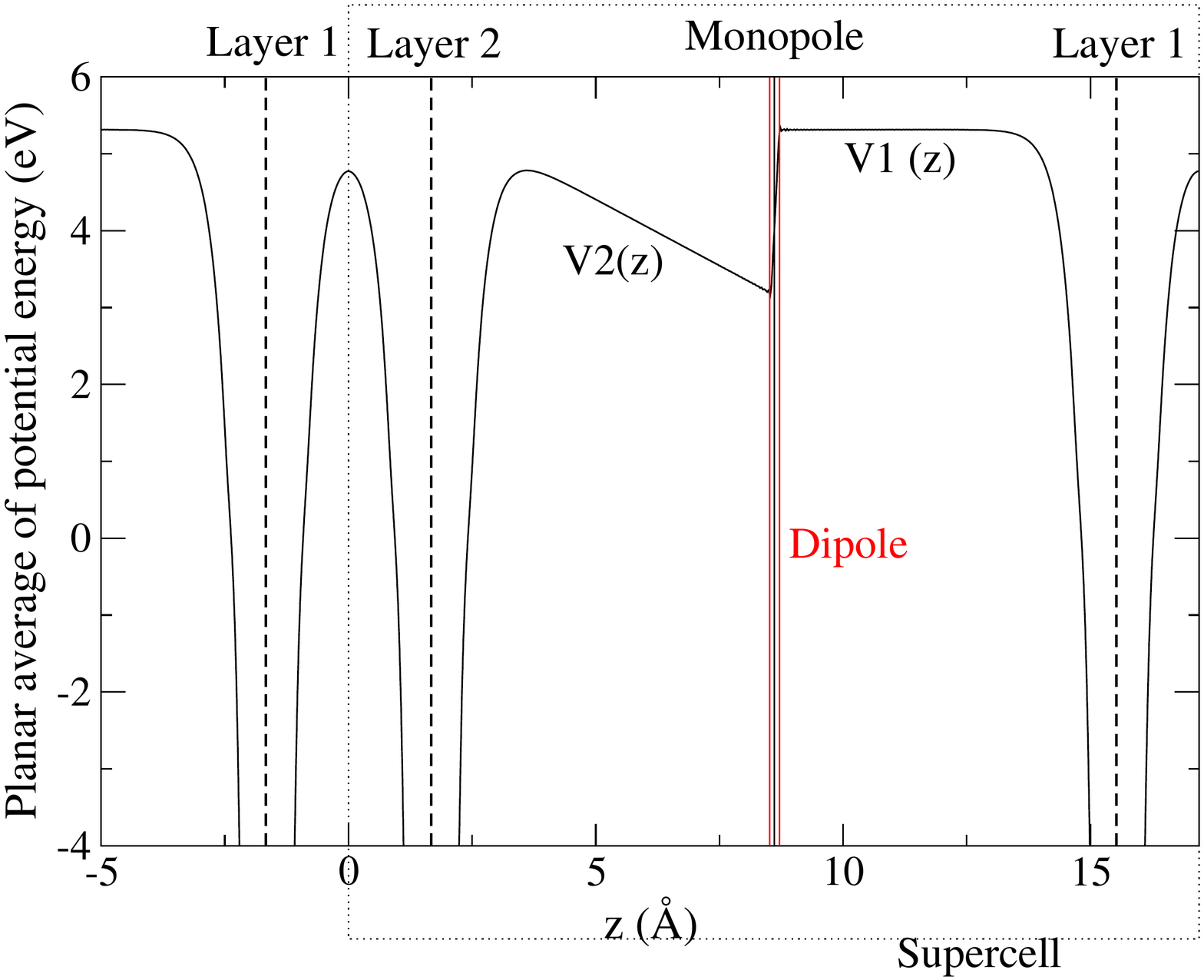}
  \caption{Planar average of the ionic, Hartree, monopole, and dipole potentials
     multiplied by the electron charge.
The positions of the first and second layers of the bilayer, of the monopole, and of the dipole
in the supercell are indicated.
This figure corresponds to a doping charge on the bilayer $n$ = \cf{19\times 10^{12}\  cm^{-2}}.
The dipole potential is such that layer 1 of the bilayer
does not feel any external electric field.}
  \label{fig:pot-dip}
\end{figure}

In order to have different
bottom and top gates, we add to the monopole a sawlike potential, called dipole potential,\cite{dipole}
generated by two planes of opposite charge, as implemented in standard distributions of
the PWSCF code \cite{PWSCF_WEB}.
The dipole is centered around the monopole, and the distance between the dipole planes
is kept fixed to 0.17 \cf{\AA}.
In Fig.~\ref{fig:pot-dip} we show the planar average of the ionic, Hartree, monopole, and
dipole potentials multiplied by the electron charge
for a doping charge $n$ = \cf{19\times 10^{12}\  cm^{-2}}. In the case shown in the figure,
the dipole potential is chosen to create a flat potential
and zero electric field on layer 1 of the bilayer.
This configuration corresponds to the case where only a bottom gate acts on the bilayer.
By changing the sign to the dipole
potential, we can obtain the opposite configuration,
with a flat potential and zero electric field on layer 2 of the bilayer.

The electric fields $E_1$ and $E_2$ are calculated from the planar average of the
ionic, Hartree, monopole, and dipole potential energy $V_1(z)$ and $V_2(z)$ on side 1 and side 2
of the bilayer, respectively, as
\begin{eqnarray}
E_1=-\left(\frac{1}{e}\right)\frac{dV_1(z)}{dz}, \\
E_2=-\left(\frac{1}{e}\right)\frac{dV_2(z)}{dz}.
\end{eqnarray}
In order to deal with uniform $E_1$ and $E_2$ electric fields, these derivatives are
calculated in the linear part of $V_1(z)$ and $V_2(z)$ (see Fig.\ref{fig:pot-dip}).

Varying independently the dipole potential and the total charge on the sample
and monopole,
one can explore all the situations with different doping $n$ on the bilayer and
different $E_{av} = (E_1 + E_2)/2$.


\newpage
\bibliography{biblio.bib}

\begin{thebibliography}{29}
\bibitem[{not()}]{note_pt}
\bibinfo{note}{This selection rule is strictly valid for the present TB
  Hamiltonian, which has only the out-of-plane hopping parameter between $A$
  and $A'$ atoms. In this case, the eigenvectors 1 and 3 of
  $\hat{H}^{0}_{\bf{k}}$ are four-component vectors of the form
  $(1,a,-1,-a^*)$. Eigenvectors 2 and 4 are of the form $(1,a,1,a^*)$. Because
  of this, given the definition of $\widehat{\Delta \rho}$
  [Eq.(\ref{pippoequation})], it can be easily shown that $\eta_{1,3}=0$ and
  $\eta_{2,4}=0$. For the most general Hamiltonian, $\eta_{1,3}$ and
  $\eta_{2,4}$ are not exactly zero but are still much smaller than the other
  transitions.}
\bibitem[{par()}]{parametro}
\bibinfo{note}{In Refs. \cite{Castroneto_PRL} and \cite{Castroneto_arXiv} the
  authors compared the experimental results on the cyclotron mass with TB
  calculations, obtained using a parameter $\gamma_{\perp} = 0.22$~eV. This
  value of $\gamma_{\perp}$ is chosen so that theory and experiment give the
  same result for the cyclotron mass at $n$ = 3.6 $\times$ \cf{10^{12}
  cm^{-2}}. However, $\gamma_{\perp}$ corresponds to the band splitting at $K$
  between the lowest occupied $\pi$ band and the highest unoccupied $\pi$ band
  (see Fig.\ref{bande}). $\gamma_{\perp}$ can be measured with different
  experimental techniques,~\cite{Ohta_science,Malard-PRB,Kuzmenko-infrared} and
  it is found to be around $0.4$~eV. Therefore, here we compare the
  experimental results with the cyclotron mass calculated in the TB formalism
  as described in Sec. \ref{sec:com}, using $\gamma_{\perp} = 0.4$~eV, in
  agreement with values found in literature.}
\end{thebibliography}

\end{document}